\documentclass[a4paper,fleqn,usenatbib]{mnras}

\usepackage[T1]{fontenc}
\usepackage{ae,aecompl}

\usepackage{graphicx}	
\usepackage{amsmath}	
\usepackage{amssymb}	
\usepackage{threeparttable}
\usepackage{comment}

\def\chandra    {{\em Chandra}\/}
\def\cha    {{\em Chandra}\/}

\def\rosat      {{\em ROSAT}\/}

\def\ga         {{CGCG~254-021}\/}

\title[Two spectacular X-ray tails]{A detached double X-ray tail in the merging galaxy cluster Z8338 with a large double tail}

\author[Ge et al.]{Chong Ge$^{1}$\thanks{E-mail: chongge@xmu.edu.cn},
Ming Sun$^{2}$\thanks{E-mail: ming.sun@uah.edu},
Paul E. J. Nulsen$^{3,4}$,
Craig Sarazin$^{5}$,
Maxim Markevitch$^{6}$,
\newauthor
and Gerrit Schellenberger$^{3}$ \\
$^{1}$Department of Astronomy, Xiamen University, Xiamen, Fujian 361005, China\\
$^{2}$Physics Department, University of Alabama in Huntsville, Huntsville, AL 35899, USA\\
$^{3}$Center for Astrophysics | Harvard and Smithsonian, 60 Garden Street, Cambridge, MA 02138, USA\\
$^{4}$ICRAR, University of Western Australia, 35 Stirling Hwy, Crawley, WA 6009, Australia\\
$^{5}$Department of Astronomy, University of Virginia, PO Box 400325, Charlottesville, VA 22904, USA\\
$^{6}$NASA/Goddard Space Flight Center, Greenbelt, MD 20771, USA
}

\date{Accepted XXX. Received YYY; in original from ZZZ}

\pubyear{2023}

\begin{document}
\label{firstpage}
\pagerange{\pageref{firstpage}--\pageref{lastpage}}
\maketitle

\begin{abstract}
When subhalos infall into galaxy clusters, their gas content is ram pressure stripped by the intracluster medium (ICM) and may turn into cometary tails.
We report the discovery of two spectacular X-ray double tails in a single galaxy cluster, Z8338, revealed by 70 ks \chandra\ observations.
The brighter one, with an X-ray bolometric luminosity of $3.9 \times 10^{42}{\rm\ erg\ s}^{-1}$, is a detached tail stripped from the host halo and extended at least 250 kpc in projection. The head of the detached tail is a cool core with the front tip of the cold front $\sim$ 30 kpc away from the nucleus of its former host galaxy. The cooling time of the detached cool core is $\sim 0.3$ Gyr.
For the detached gas, the gravity of the once-associated dark matter halo further enhances the Rayleigh-Taylor (RT) instability. 
From its survival, we find that a magnetic field of a few $\mu$G is required to suppress the hydrodynamic instability.
The X-ray temperature in the tail increases from 0.9 keV at the front tip to 1.6 keV in the wake region, 
which suggests the turbulent mixing with the hotter ICM.
The fainter double X-ray tail, with a total X-ray luminosity of $2.7 \times 10^{42}{\rm\ erg\ s}^{-1}$, appears to stem from the cool core of a subcluster in Z8338, and likely was formed during the ongoing merger. 
This example suggests that X-ray cool cores can be displaced and eventually destroyed by mergers, while the displaced cool cores can survive for some extended period of time.
\end{abstract}




\section{Introduction}

The hierarchical structure formation theory posits that large halos are formed through subhalo mergers. While the merger history of dark matter halos can be tracked well with N-body simulations, it is the baryon physics that limits our understanding of galaxy formation and cluster evolution.
One important element of baryon physics is ram pressure stripping (RPS; \citealt{Gunn1972}), which determines the gas content of subhalos (see the recent review by \citealt{2022A&ARv..30....3B}). Galaxy formation models generally rely on hot halo gas to feed the growth of massive galaxies \citep[e.g.][]{WF91,Bower06}.
The early models assumed an instantaneously complete gas removal when galaxies enter groups or clusters \citep[e.g.][]{WF91}. This is now known to be too simplistic, as \chandra\ observations have revealed that hot gas halos can survive in galaxies of groups and clusters \citep[e.g.][]{Sun2007,Jeltema08,Sun2009b}. The desire to have a more realistic treatment of RPS in galaxy evolution has triggered many studies \citep[e.g.][]{Font08,McCarthy08,Steinhauser2016}.
While these new implementations do seem to alleviate some early problems, e.g. the relative fraction of blue and red galaxies, the RPS prescription used in cosmological simulations and semi-analytic models is still too simple and needs to be calibrated with more observational data and detailed simulations.

Galaxy clusters provide great labs to study RPS, as subhalos
(galaxies or subclusters) soar through the hot intracluster medium (ICM) all the time.
\chandra\ has revolutionized this field, with detections of cold fronts and shocks \citep[e.g.][]{markevitch2007}.
While many examples of RPS have been found \citep[e.g.][]{Machacek06,Randall08,Russell14,Sun2022}, very few detached hot halos have been detected, especially high-density ones.
In fact, the systematic analysis of \cite{Sun2007} detected no detached tails in 25 clusters. M86's tail has a prominent detached component, but M86 still hosts a dense X-ray cool core (\citealt{Randall08}).
Detached hot halos are observed in simulations \citep[e.g.][]{McCarthy08,Vij15}.

It is important to study detached hot halos, especially high-density ones.
First, high-density detached hot halos can be considered detached cool cores (CCs).
They are not gravitationally bound but cooling time may still be less than 1 Gyr. They are great targets to study energy transfer to compare with bound CCs.
Second, detached hot halos are great targets to study hydrodynamic instability.
As the infalling dense cloud is decelerating by the ram pressure of ambient ICM, there is an inertial force directed from the dense phase to the less-dense phase, so the infalling cloud should suffer from the Rayleigh-Taylor (RT) instability.
Typically, infalling clouds are still bound so the gravity of the associated dark matter (DM) halo helps to suppress RT instability \citep{markevitch2007}, while the self-gravity of the X-ray gas is usually too small to have any impact.
In principle, without the suppression of the RT instability by the gravity of DM halos, detached hot halos should disintegrate quickly, as shown in simulations \citep[e.g.][]{Jones96}.
The Bullet cluster does host a detached cool core, 140 kpc from the once associated
dark matter halo, but the detached core in A520 is believed to 
have been destroyed by RT instability \citep[e.g.][]{markevitch2007}.
Third, detached hot halos will contribute to the clumping of the ICM so their survival and evolution are important for understanding the clumps in the ICM \citep[e.g.][]{Vazza2013,Ge2021a}.

\begin{figure}
\begin{center}
\centering
\includegraphics[angle=0,width=0.49\textwidth]{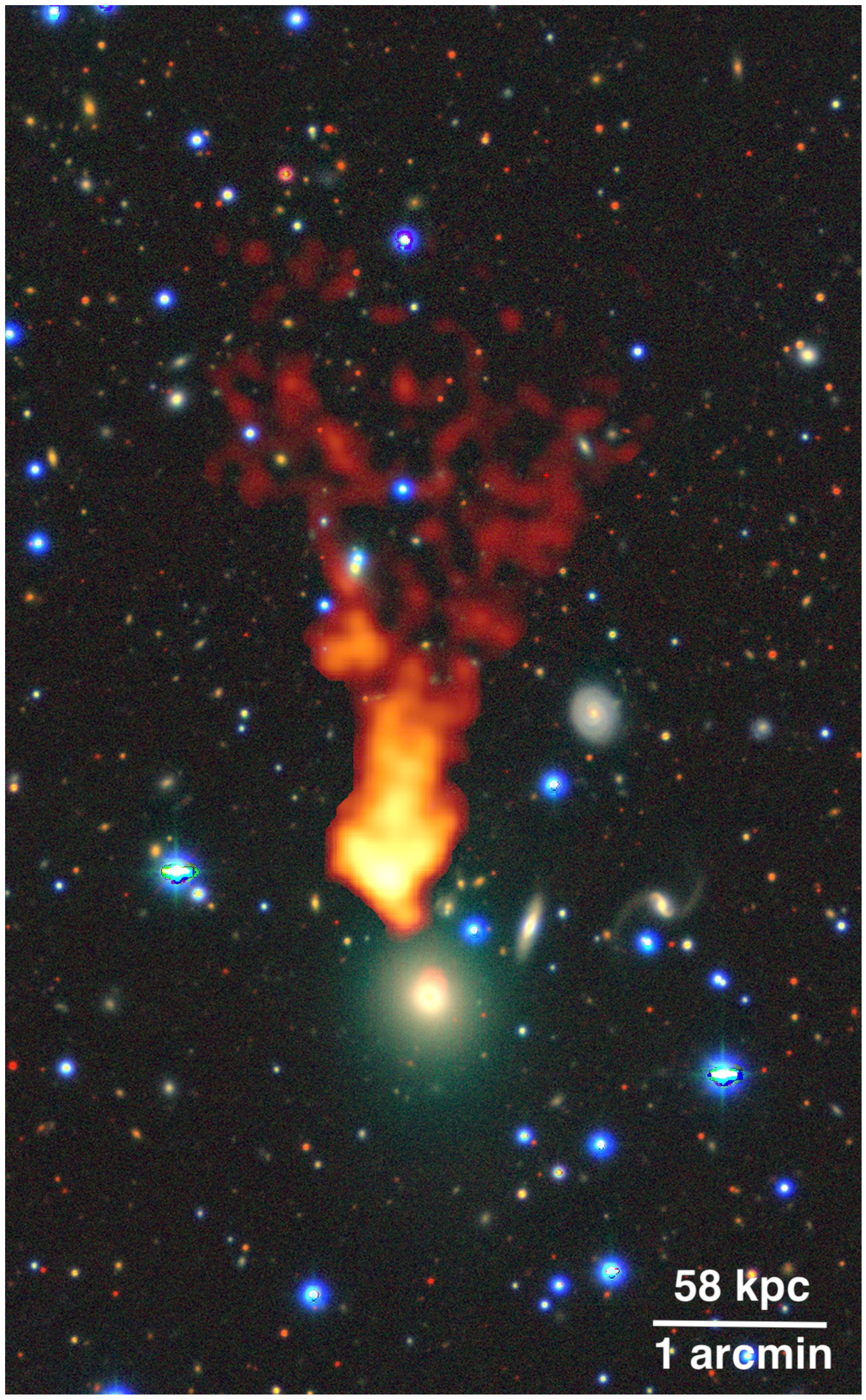}
\vspace{-0.5cm}
\caption{
The composite optical/X-ray image of the detached tail and its former host galaxy CGCG 254-021. The background optical image is from DESI Legacy Imaging Surveys (e.g. \citealt{Dey2019}), overlaid with the warm color \cha\ 0.7-2 keV image. 
The \cha\ image is cropped around the tail to highlight it, images with a larger coverage are shown in Fig.~\ref{fig:img}.
As a subhalo infalling into the cluster, its gas content is ram pressure stripped by the ICM. In this case, the hot halo is stripped and detached from the host galaxy.
Note in the galaxy center, a remnant faint X-ray corona survives but also offsets toward the north. 
}
\label{fig:rgb}
\end{center}
\end{figure}

MCXC J1811.0+4954 \citep{Piffaretti2011} or ZwCl8338 (hereafter Z8338) is a poor cluster with a system temperature of $\sim 3$ keV. 
Z8338 hosts three brightest cluster galaxies (BCGs) with comparable near-infrared luminosity (as a proxy of the stellar mass; Table~\ref{Gs}). The global X-ray morphology in the {\em ROSAT} image shows elongated and disturbed features with two X-ray peaks. The optical and X-ray data indicate that Z8338 is a dynamic young cluster being assembled.
The center of Z8338 is occupied by a pair of bright elliptical galaxies, NGC 6582 NED01 and NGC 6582 NED02, with essentially the same radial velocities (Table~\ref{Gs}).
The two galaxies are separated by $\sim$ 30 kpc in projection.
A detached X-ray tail was studied in \cite{Schellenberger15} with 8 ks of \cha\ observation.
Its X-ray peak is not on a cluster galaxy. A large cluster galaxy, CGCG 254-021, lies nearby, but still $\sim 40$ kpc away in projection (Fig.~\ref{fig:rgb}).
Here we present new results by adding much deeper \chandra\ observations from our own program (proposal ID: 17800628, PI: Sun).
Section~\ref{sec:data} presents the \cha\ data reduction. Section~\ref{sec:result} reports the X-ray properties of the cluster and its substructures. Section~\ref{sec:discussion} is the discussion. We include our conclusions in Section~\ref{sec:conclusion}.
We assume a cosmology with $H_0$ = 70 km s$^{-1}$ Mpc$^{-1}$, $\Omega_m=0.3$, and $\Omega_{\Lambda}= 0.7$. At Z8338's redshift of $z=0.0494$ \citep{Cava2009}, $1^{\prime\prime}=0.966$ kpc.

\begin{table*}
\centering
\caption{The three brightest galaxies in the Z8338 cluster}
\label{tab_galaxies}
\begin{tabular}{ccccccc}
\hline
ID & Name & RA, DEC & velocity & W1$^{\rm a}$ & $L_{\rm 1.4 GHz}$$^{\rm b}$ & comment \\
   &      &         & (km/s)   & (mag) & ($10^{22}$ W/Hz) &       \\
\hline
G1 & NGC 6582 NED02 & 18:11:05.1, 49:54:33 & 14474$\pm$21$^{\rm c}$ & 11.083 & $< 2.7$ & associated with a small corona \\
G2 & NGC 6582 NED01 & 18:11:01.9, 49:54:43 & 14431$\pm$63$^{\rm d}$ & 11.584 & 9.8 & associated with the large cool core \\
G3 & CGCG 254-021 & 18:10:29.2, 49:55:17 & 15354$\pm$19$^{\rm c}$ & 11.583 & 9.7 & associated with the detached tail \\
\hline
\end{tabular}
\begin{flushleft}
$^{\rm a}$: {\em WISE} band 1 (3.4 $\mu$m) magnitude as a proxy of the stellar mass \\
$^{\rm b}$: 1.4 GHz luminosity from {\em NVSS} assuming a spectral index of -0.8. An upper limit of 5 mJy is assumed for G1. \\
$^{\rm c}$: \cite{Smith04}; $^{\rm d}$: \cite{1980ApJS...44..137K} \\
\end{flushleft}
\label{Gs}
\end{table*}

\section{{\em CHANDRA} analysis}
\label{sec:data}

\begin{figure*}
\begin{center}
\centering
\includegraphics[angle=0,width=0.49\textwidth]{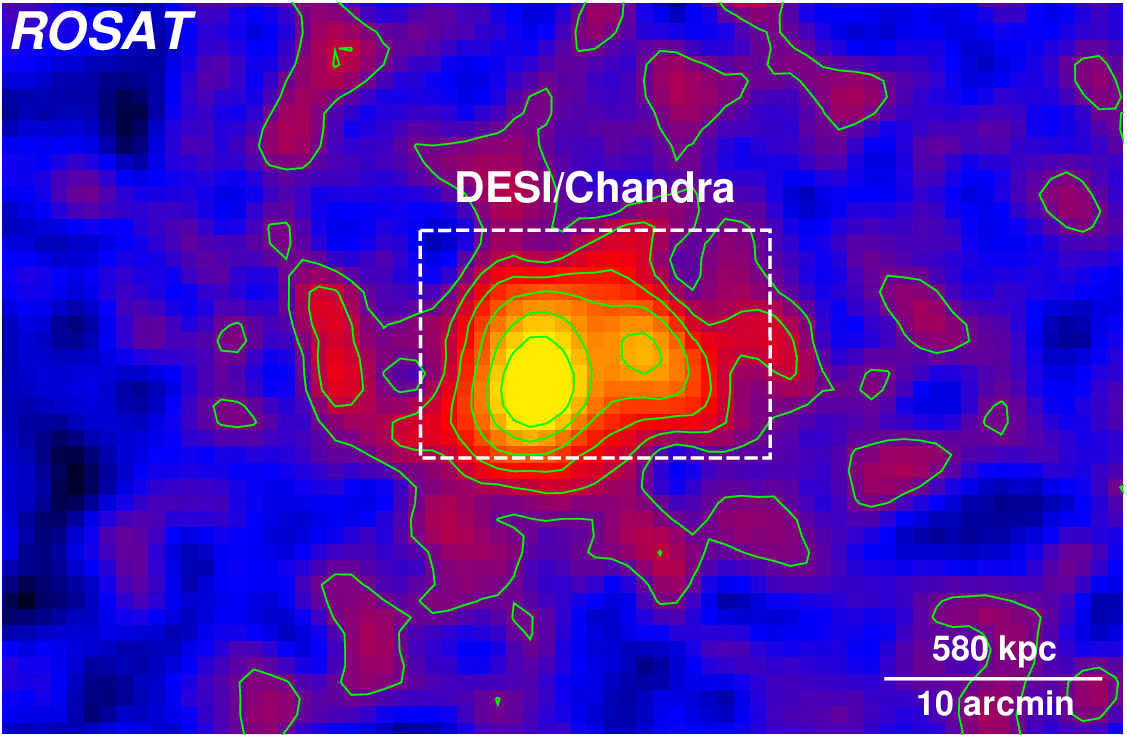}
\includegraphics[angle=0,width=0.49\textwidth]{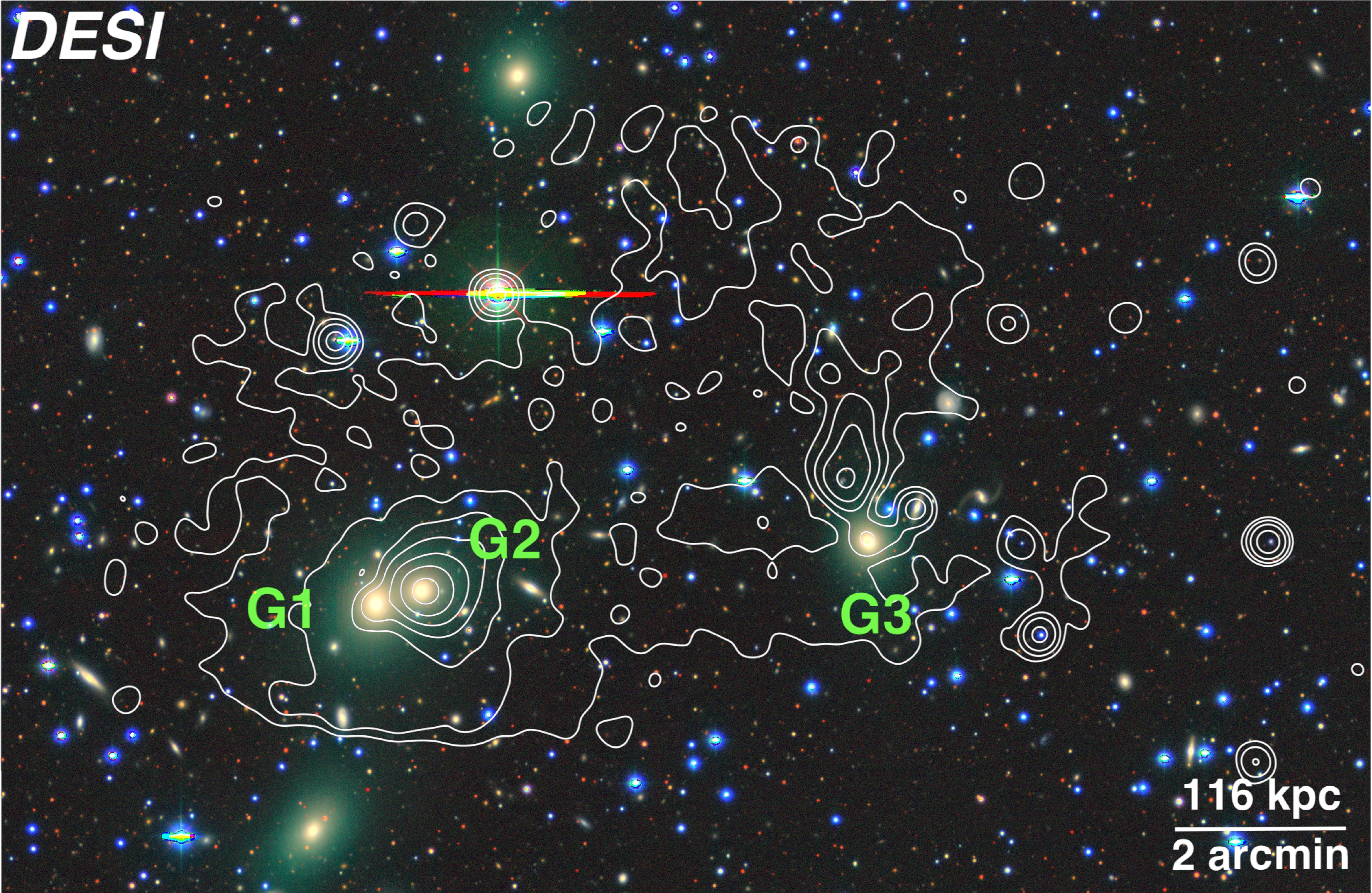}
\includegraphics[angle=0,width=0.49\textwidth]{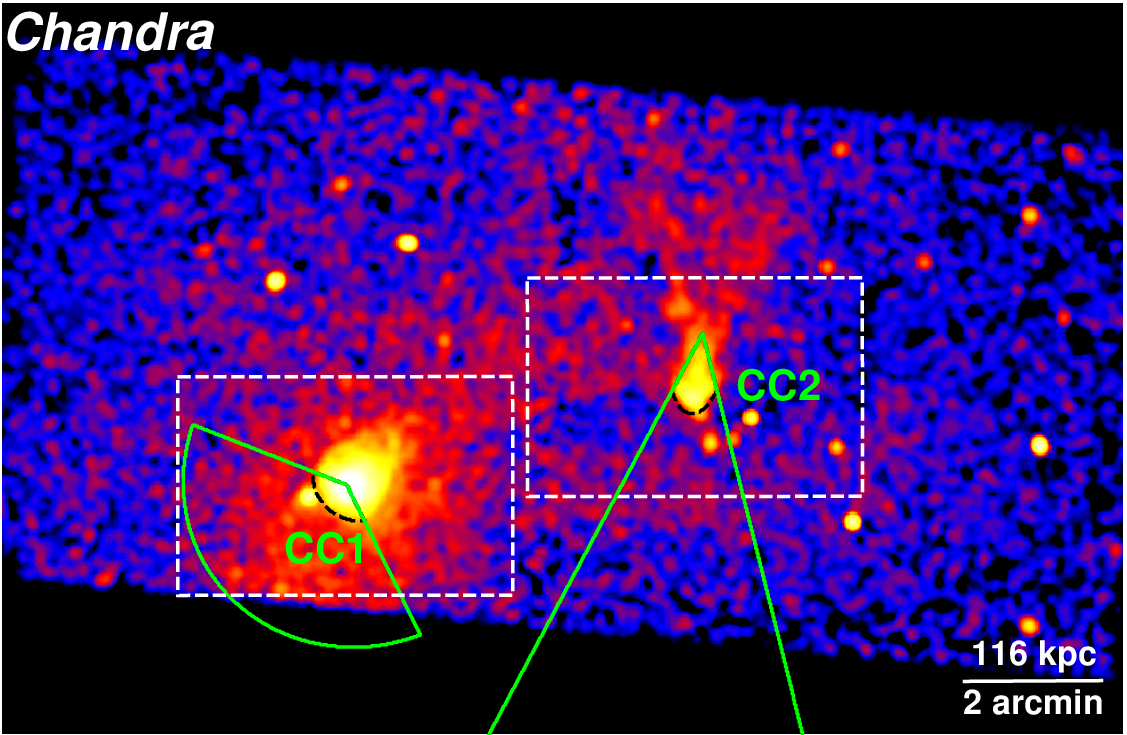}
\includegraphics[angle=0,width=0.49\textwidth]{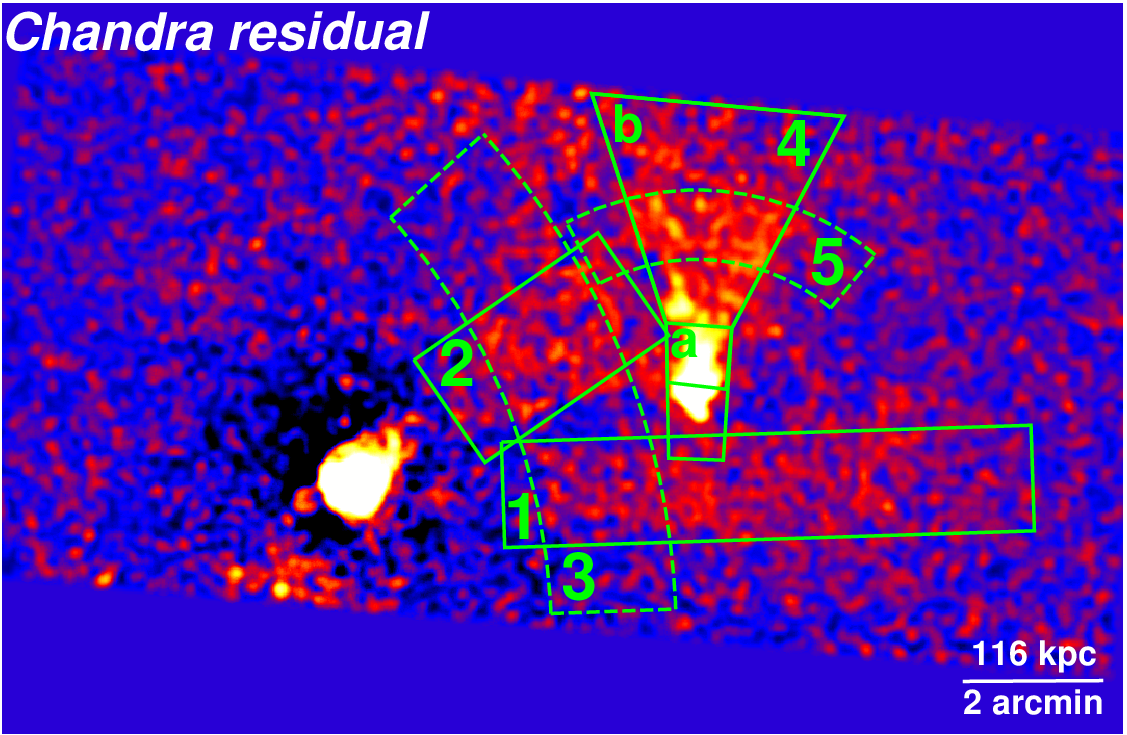}
\includegraphics[angle=0,width=0.49\textwidth]{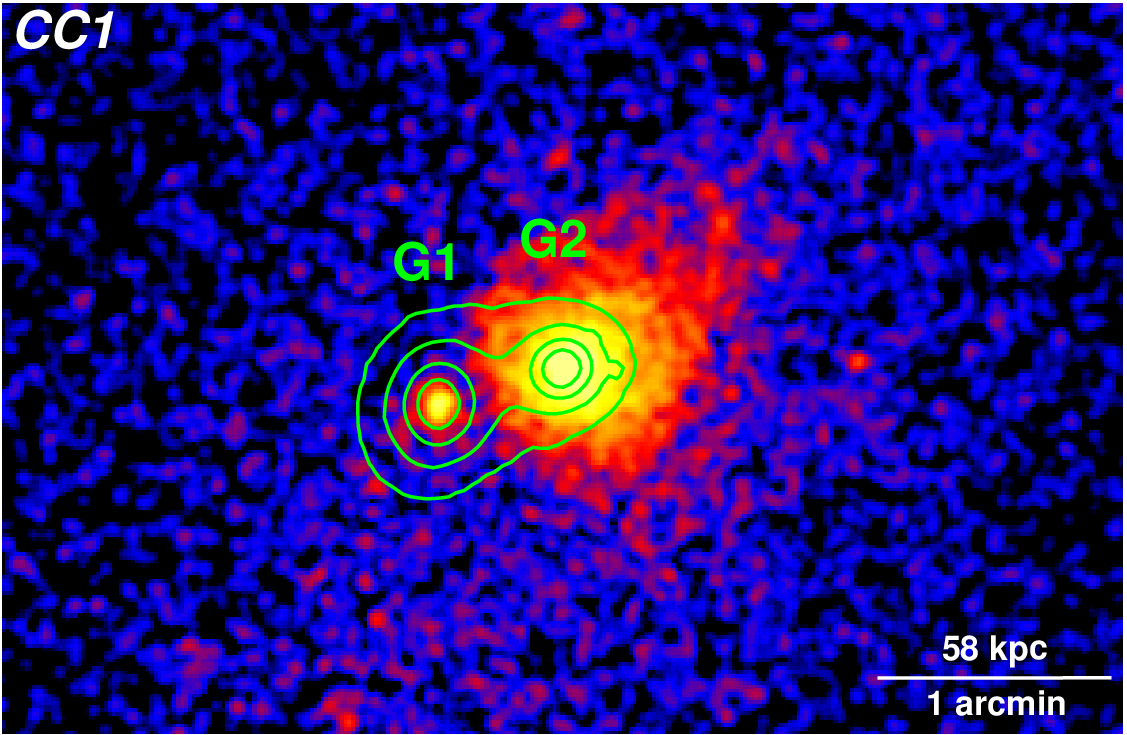}
\includegraphics[angle=0,width=0.49\textwidth]{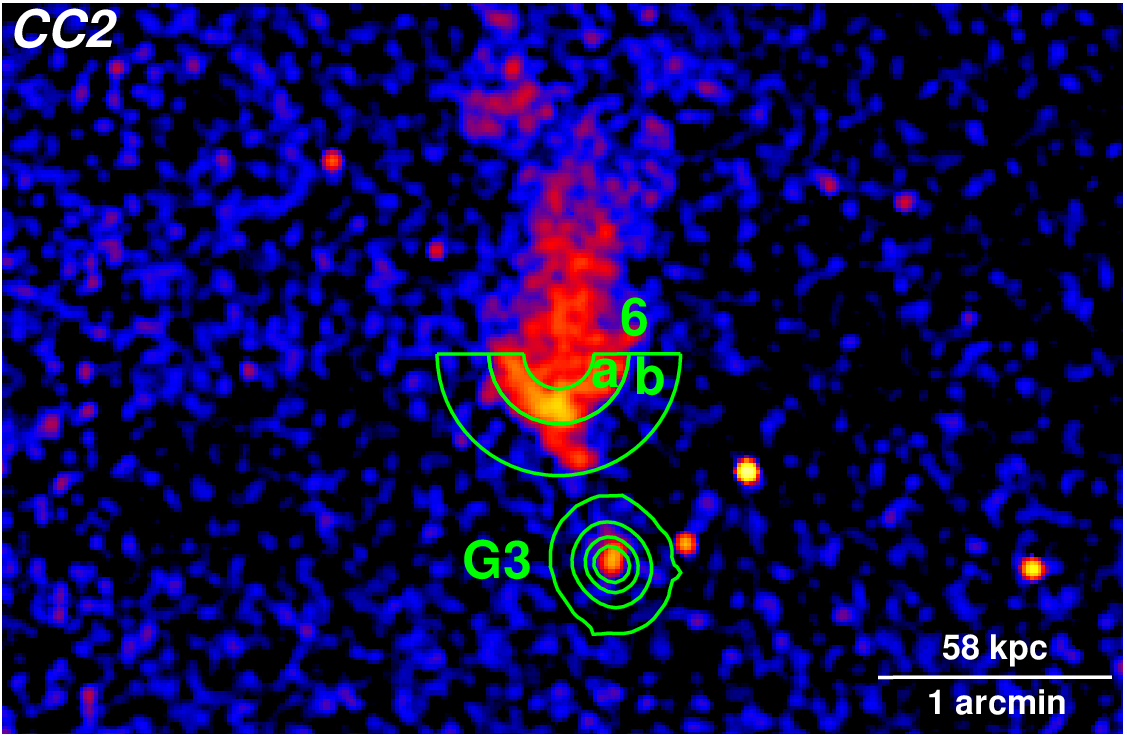}

\vspace{-0.1cm}
\caption{
{\it Top left}: 0.1-2.4 keV image for Z8338 from the \rosat\ All Sky Survey. 
The dashed box shows the region covered by the DESI Legacy Imaging Surveys in the top right panel, and by \cha\ in the middle panels.
{\it Top right}: Optical image from the DESI Legacy Imaging Surveys overlaid with contours from the \cha\ image. Three bright cluster galaxies are marked.
{\it Middle left}: background subtracted and exposure corrected 0.7-2 keV \cha\ image. The two dashed boxes outline the regions around CC1 and CC2 in the bottom panels. The green sectors are for the SBP extraction in Fig.~\ref{fig:sbp}. The black dashed arcs mark the location of cold fronts (CFs).
{\it Middle right}: \cha\ residual image excluding point sources and large-scale cluster emission. Marked regions are for extracting SBPs for tails in Fig.~\ref{fig:sbp}.
{\it Bottom left}: \cha\ image of CC1 overlaid with optical contours from WINGS $V$-band image. The galaxy G1 is associated with the small corona, while G2 is associated with the large CC1.
{\it Bottom right}: \cha\ image of CC2 overlaid with contours from the WINGS $V$-band image. G3 is associated with the detached tail and has a remnant corona at its center.
}
\label{fig:img}
\end{center}
\end{figure*}

There were three \chandra\ observations with X-ray tails covered (Table \ref{tab_obs}),
all taken with the Advanced CCD Imaging Spectrometer (ACIS). 
We analyzed the \cha\ observations with the \cha\ Interactive Analysis of Observation (CIAO; version 4.11) and Calibration Database (CALDB; version 4.8.2), following the procedures of \cite{2018MNRAS.481.4111G}.
We used {\sc chandra\_repro} script with VFAINT mode correction to produce a new level 2 event file. We applied the {\sc deflare} script to filter flare intervals. Point sources were detected with the {\sc wavdetect} script. The standard stowed background (2009-09-21 data set) with the VFAINT mode cleaning applied was matched to each observation. We used {\sc merge\_obs} to combine different observations. We then produced background subtracted and exposure corrected image as in Fig.~\ref{fig:img}.

We extracted the spectra separately in the BI and FI
chips with {\sc specextract}. 
Diffuse source spectra extracted from several data sets and chips (BI or FI)
were fit jointly, with the source spectral parameters (e.g. temperature, metallicity) tied and normalizations free for each data set.
We used the local background subtraction for point sources and double background subtraction  (e.g. \citealt{2021MNRAS.508L..69G}) for diffuse sources.
The double-background subtraction method subtracts the non-X-ray (detector) background and X-ray background separately. 
Briefly, we extract on-source and off-source spectra from the \cha\ data.
The detector backgrounds are extracted from the \cha\ stowed data set, and rescaled to the observations according to the 9.5-12 keV count rates. The detector spectra are then loaded into {\sc xspec} as background spectra. The X-ray background, including the sky and ICM components, are properly modeled in {\sc xspec}. The on-source and off-source X-ray backgrounds are linked by the best-fit ICM/sky models and sky solid angles.
This double-background subtraction method accounts for the variation of the X-ray effective area and ICM emission across the field, as well as the potential change of the non-X-ray background across the detector. 
The solar photospheric abundance table of \cite{Asplund09}
was used in the spectral fits. Uncertainties quoted in this paper are 1$\sigma$.
In the spectral analysis, the Cash statistics in {\sc xspec} (cstat) was used.
We adopted an absorption column density of 4.77$\times10^{20}$ cm$^{-2}$, obtained using the tool\footnote{http://www.swift.ac.uk/analysis/nhtot/index.php}
of \cite{Willingale13}.
Absorption by the interstellar medium was modeled with {\sc tbabs} in {\sc xspec}.

\begin{table}
\protect\caption{\chandra\ Observations}
\begin{tabular}{cccc}
\hline
ObsID (PI) & Date Obs & Detector & Total/Clean Exp \\
      &          &          & (ks)        \\
\hline
15163 (Reiprich) & 2013-01-04 & ACIS-I & 8.1/8.0 \\
18281 (Sun) & 2016-12-27 & ACIS-S & 31.6/31.2 \\
19978 (Sun) & 2016-12-26 & ACIS-S & 30.1/29.3 \\
\hline
\end{tabular}
\label{tab_obs}
\end{table}

\begin{figure*}
\begin{center}
\centering
\includegraphics[angle=0,width=0.32\textwidth]{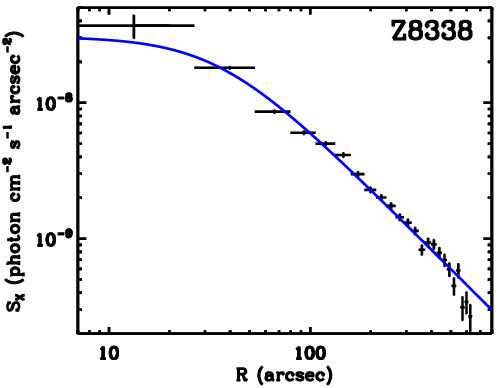}
\includegraphics[angle=0,width=0.32\textwidth]{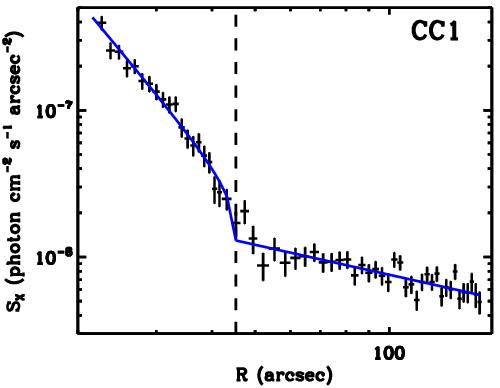}
\includegraphics[angle=0,width=0.32\textwidth]{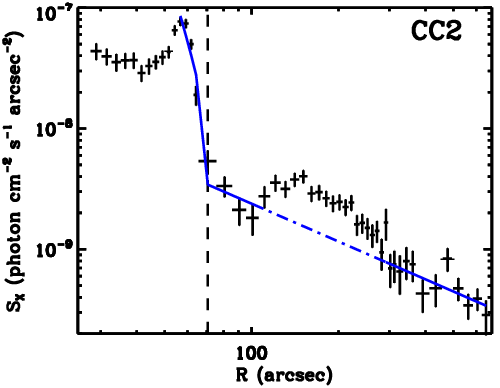}
\includegraphics[angle=0,width=0.32\textwidth]{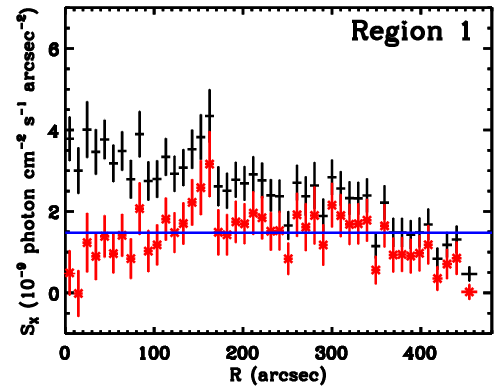}
\includegraphics[angle=0,width=0.32\textwidth]{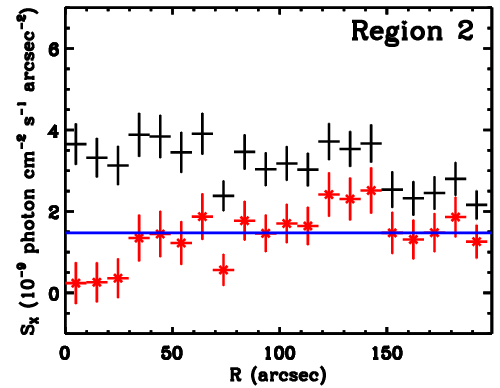}
\includegraphics[angle=0,width=0.32\textwidth]{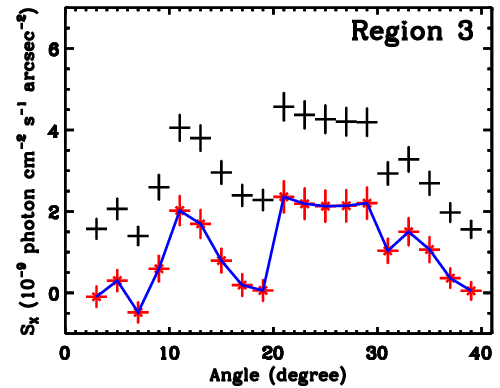}
\includegraphics[angle=0,width=0.32\textwidth]{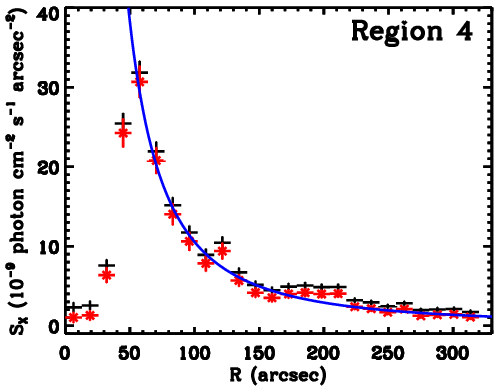}
\includegraphics[angle=0,width=0.32\textwidth]{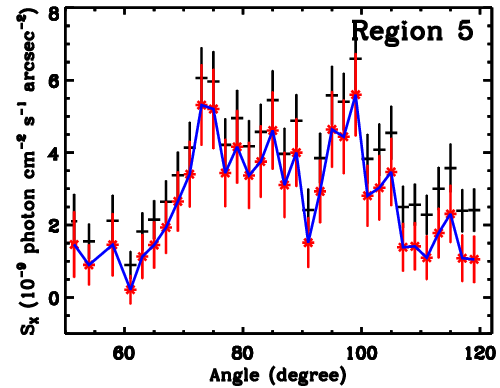}
\includegraphics[angle=0,width=0.32\textwidth]{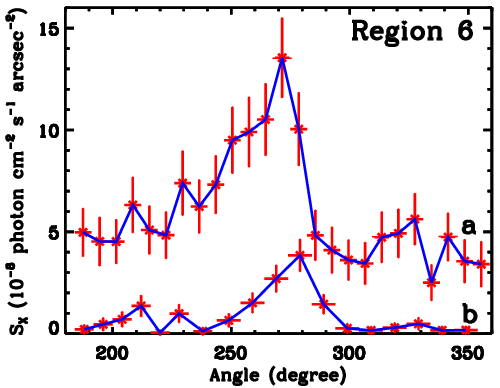}

\vspace{-0.1cm}
\caption{
SBPs from different regions in Fig.~\ref{fig:img} as marked in the upper-right corner.
{\it Z8338}: SBP of the host cluster. The solid line is the best-fit $\beta$-model.
{\it CC1 and CC2}: SBPs from the green sectors marked in the middle left panel of Fig.~\ref{fig:img}. The solid lines are the best-fit projected, broken power-law model for the density profile. The dashed lines mark the locations of the CFs. The intensity enhancement above the dash-dotted line of CC2 is from the tail structure within region 1.
{\it Regions 1-6}: SBPs for tails from marked regions in Fig.~\ref{fig:img}. The black pluses are from the original intensity image, while the red crosses are from the residual image with the large-scale cluster gradient subtracted. The SBPs from regions 1, 2, and 4 are radial profiles. The SBPs from regions 1 and 2 are nearly uniform, the solid lines are the median intensity. The SBP of the tail in region 4 can be fit with a power-law function shown as a solid line. The SBPs from regions 3, 5, and 6 are azimuthal profiles. The roughly double peak features of regions 3 and 5 highlight the double tail structures. The dominant peak in region 6 is from the protrusion.
}
\label{fig:sbp}
\end{center}
\end{figure*}

\section{results}
\label{sec:result}
Fig.~\ref{fig:img} shows X-ray and optical images of Z8338.
Significant features of Z8338, including X-ray contours, are indicated on the optical image in the top right panel of Fig. 2.  We take the center of the cluster to be located at the center of the cool core, CC1, at (RA=18:11:01.6, DEC=+49:54:42.2), which lies close to the center of the BCG, NGC 6582 NED01, labelled G2.  A much smaller X-ray corona is centered close to the second BCG, NGC 6582 NED02, labelled G1.  The detached tail is associated with a third BCG, CGCG 254-021, labelled G3 (more details are listed in Table~\ref{tab_galaxies}).

\subsection{Host cluster}
\label{sec:cluster}

We first want to derive the X-ray properties of the host cluster with the \cha\ data.
We extract a surface brightness profile (SBP) centered on CC1 from the merged intensity map (Fig~\ref{fig:img}), excluding regions of point sources and diffuse substructures (CCs and tails).
We apply a $\beta$-model \citep{1976A&A....49..137C} to fit the SBP of host ICM in Fig.~\ref{fig:sbp} (top left panel).
The $\beta$-model gas distribution is given by $n_{\rm ICM}(r)=n_{\rm e0}[1+(r/r_c)^2]^{-3\beta/2}$, which is an analytical model with the derived X-ray SBP also following  a $\beta$-model in the form of $I_{\rm X}(r)=I_0[1+(r/r_c)^2]^{1/2-3\beta}$.
We use the analytical formula Eq.~(4) of \cite{2018MNRAS.481.4111G} to convert the central surface brightness $I_0$ (from the $\beta$-model fitting to the SBP) to the central electron density $n_{\rm e0}$.
The best-fit parameters are $n_{\rm e0}=(6.1\pm0.4)\times 10^{-3}{\rm\ cm^{-3}}$, $I_0=(3.1\pm0.3)\times 10^{-8}{\rm\ photon\ cm^{-2}\ s^{-1}\ arcsec^{-2}}$, $r_c=35.1\pm3.4$ arcsec, and $\beta=0.41\pm0.01$.

We then measure the mean cluster temperature within 0.15-0.75 $R_{500}$ excluding substructures like CCs and tails, and assume a mean ICM abundance of 0.3 solar, where $R_{500}$ is from the $M-T_X$ relation \citep{Sun2009}.
The best-fit mean temperature is $3.1\pm0.2$ keV,
the related mass and radius are $M_{500}=1.9\times10^{14}M_\odot$ and $R_{500}=850{\rm\ kpc}$ or 14.7 arcmin.
Integrating $n_{\rm ICM}(r)$ within $R_{500}$ gives a total hot ICM mass $M_{g,500}=1.4\times10^{13}M_\odot$.

\subsection{Cool cores and coronae}

The BCGs in Table~\ref{tab_galaxies} are associated with the X-ray bright substructures in Fig~\ref{fig:img} (bottom panels). We extract the X-ray spectra of these substructures and list their properties in Table~\ref{t:src}. The X-ray spectral information suggests that they are either a large CC or a small corona.

CC1 is associated with G2, and CC2 lies close to G3, though detached from it.
The X-ray temperatures of CC1 and CC2 are $T_{\rm cc1}=1.7$ keV and $T_{\rm cc2}=0.88$ keV, while the temperatures of the surrounding ICM are $T_{\rm icm1}=3.1$ keV and $T_{\rm icm2}=3.5$ keV, respectively.
Both CCs are cooler than the ambient ICM, and are accompanied by sharp surface brightness edges known as cold fronts (CFs), which are marked as dashed arcs in Fig.~\ref{fig:img} (middle left panel). 
We extract SBPs from elliptical annuli within the sectors marked in the middle left panel of Fig.~\ref{fig:img} and then fit the SBPs with broken power-law functions for the CFs in Fig.~\ref{fig:sbp}.
The derived gas density jumps of CF1 and CF2 are $\rho_{\rm j,cf1}=2.6 \pm 0.2$ and $\rho_{\rm j,cf2}=3.9 \pm 0.7$.

We then estimate the mean gas density inside the CCs, from the {\sc apec} normalization:
\begin{equation}
\eta=\frac{10^{-14}}{4\pi [D_A(1+z)]^2}\int n_e n_H dV,
\label{eq:eta}
\end{equation}
where the angular size distance $D_A=199.3$ Mpc (all the length units are converted to cm when estimating the density) at the redshift $z=0.0494$.
We assume a spherical cloud ($r=27''$) and a hemispherical 
cloud ($r=17''$) of uniform density for CC1 and CC2, respectively.
The emission measure is $\int n_e n_H dV=n_e^2\frac{n_H}{n_e}\frac{4\pi}{3} r^3$ for CC1, halved for CC2. 
Using the {\sc xspec} norms of $\eta=(1.0\pm 0.1) \times 10^{-3}$ for CC1 and $\eta=(4.7\pm 2.0) \times 10^{-5}$ for CC2, 
the resultant mean densities are $n_{\rm e,cc1}=(1.7\pm 0.1)\times 10^{-2}{\rm\ cm^{-3}}$ and $n_{\rm e,cc2}=(1.0\pm 0.2)\times 10^{-2}{\rm\ cm^{-3}}$.
The hot gas masses of CCs are $M_{\rm g,cc1}=(3.6\pm0.2) \times 10^{10} M_{\odot}$ and $M_{\rm g,cc2}=(2.6\pm0.2) \times 10^{9} M_{\odot}$ for CC1 and CC2, respectively.
We use the density with the temperature and metallicity from spectral fitting to estimate the cooling times of CCs.
The cooling time is estimated by the total thermal energy divided by the X-ray emissivity, or $3/2 n_{\rm total} kT / (n_{\rm e} n_{\rm H} \Lambda$)
where $\Lambda$ is the cooling rate calculated from the {\sc apec} model in {\sc xspec}, in the 0.01 - 100 keV band.
The cooling times of CC1 and CC2 are $0.95\pm0.13$ Gyr and $0.34\pm0.25$ Gyr.
For CC1, if we apply the analytical $\beta$-model analysis as in Sec.~\ref{sec:cluster}, we estimate a central density of $n_{\rm e0,cc1}=(6.3\pm 0.3)\times 10^{-2}{\rm\ cm^{-3}}$.
The resultant central cooling time is $0.26\pm0.04$ Gyr.

The compact sources over the nuclei of G1 and G3 are thermal coronae, based on two criteria \citep{Sun2007}:
1) their images are bright and extended in the soft band (0.7-2 keV) and faint in the hard band (2-7 keV); 2) their spectra are each poorly fit by a power-law model and have a significant iron L-shell hump at $\sim 0.9$ keV. The best-fit temperatures for thermal coronae of G1 and G3 are 1.38 keV and 0.74 keV, respectively.
The radii of coronae are 5.4$''$ and 3.7$''$ for G1 and G3. Assuming a constant density for the coronae, the average densities are $n_{\rm e,g1}=(3.7 \pm 0.3)\times 10^{-2}$ cm$^{-3}$ and $n_{\rm e,g3}=(2.8 \pm 0.6)\times 10^{-2}$ cm$^{-3}$.
The X-ray gas masses of coronae are $M_{\rm g,g1}=(6.3\pm0.6) \times 10^{8} M_{\odot}$ and $M_{\rm g,g3}=(1.5\pm0.3) \times 10^{8} M_{\odot}$.
The cooling times of the G1 and G3 coronae are $0.28\pm0.06$ Gyr and $0.13\pm0.06$ Gyr.

\begin{table}
 \centering
  \caption{X-ray properties of CCs and coronae}
\tabcolsep=0.15cm  
  \begin{tabular}{@{}lccccccc@{}}
\hline
Source$^a$ & $kT^b$ & $Z^b$  & $L_{\rm bol}^c$ & Cstat/DOF\\
& (keV) & (solar) & (erg s$^{-1}$) & \\
\hline
CC1 & $1.70\pm0.05$ & $0.65\pm0.12$ & $(8.7\pm1.2)\times 10^{42}$ & 174/161 \\
CC2 & $0.88\pm0.03$ & $1.11\pm0.53$ & $(7.4\pm3.1)\times 10^{41}$ & 25/23 \\
G1 &  $1.38\pm0.10$ & $0.8$  & ($4.4\pm0.9) \times 10^{41}$ &  13/12\\
G3 & $0.74\pm0.16$ & $0.8$  & $(1.5\pm0.5) \times 10^{41}$   & 10/8\\
\hline
\end{tabular}
\begin{tablenotes}
\item
$^a:$ The CCs and coronae are shown in Fig.~\ref{fig:img}.
$^b:$ The best-fit temperature and abundance of the {\sc apec} model.
For G1 and G3, the abundance could not be well constrained by the spectrum, and we fixed it at a typical value for galactic coronae.
$^c:$ The total X-ray bolometric luminosity.
\end{tablenotes}
\label{t:src}
\end{table}

\begin{table}
 \centering
  \caption{X-ray properties of tails}
\tabcolsep=0.15cm  
  \begin{tabular}{@{}lccccccc@{}}
\hline
Source$^a$ & $kT^b$ & $Z^b$  & $L_{\rm bol}^c$ & Cstat/DOF\\
& (keV) & (solar) & (erg s$^{-1}$) & \\
\hline
Reg1 & $1.01\pm0.04$ & $0.72\pm0.27$ & $(1.4\pm0.5)\times 10^{42}$ & 402/420 \\
Reg2  & $4.44\pm1.13$ & $0.3$ & $(1.3\pm0.1)\times 10^{42}$ &  375/387 \\
Reg1 (T) & $1.24\pm0.04$ & $0.3$ & $(3.1\pm0.1)\times 10^{42}$ & 102/88 \\
Reg2 (T)  & $3.98\pm0.38$ & $0.3$ & $(3.4\pm0.2)\times 10^{42}$ &  54/54 \\
Reg4a  & $0.95\pm0.04$ & $0.23\pm0.04$ & $(1.4\pm0.2)\times 10^{42}$ & 348/353 \\
Reg4b  & $1.58\pm0.11$ & $0.27\pm0.07$ & $(2.5\pm0.3)\times 10^{42}$ & 394/412 \\
\hline
\end{tabular}
\begin{tablenotes}
\item
The spectral properties of Reg1 and Reg2 are from the double-background subtraction method.
We also use the naive single $T$ model to fit the spectra in Reg1 and Reg2 (3rd and 4th rows), assuming the X-ray emission is dominated by the tails. 
For region 4, the CC2 head is masked, Reg4a represents the remnant tail, and Reg4b for the wake.
\end{tablenotes}
\label{t:tail}
\end{table}

\subsection{Tails}
As shown in Fig.~\ref{fig:img} (middle left panel), diffuse tails trail behind both CCs.
To enhance the tail features, the large-scale average cluster emission represented by the best-fit $\beta-$model in Section~\ref{sec:cluster} is subtracted from the original intensity map.
The residual image is shown in Fig.~\ref{fig:img} (middle right panel). 
We then extracted SBPs from both original and residual intensity maps within regions marked in Fig.~\ref{fig:img}, and plot the SBPs in Fig.~\ref{fig:sbp}. 
Regions 1 and 2 enclose double tails behind CC1.
Region 3 contains tails and nearby background for highlighting the double tail structure.
Region 4 includes the detached tail behind CC2. It is further divided into regions a (remnant tail) and b (wake). Region 5 is also used to contrast the double tail structure.
Region 6 is used for azimuthal profiles near the CC2, especially for the protrusion or finger-like part pointing to the remnant corona. 
From the images and SBPs, both tails show double tail structures, which are highlighted by the double peak features in the azimuthal SBPs from the arc shaped regions 3 and 5 across tails. 
We also extract the X-ray spectra from tail structures and list their properties in Table~\ref{t:tail}. The overall X-ray properties (such as $T_X$ and $L_X$) of the detached tail are consistent with the one reported in \cite{Schellenberger15}. Our deeper observations reveal more detailed features as presented below.

Regions 1 and 2 extend along the two tails associated with CC1. Their SBPs in Fig.~\ref{fig:sbp} indicate that the X-ray intensity is roughly uniform along the tails, and the solid line is the median intensity.
We extract X-ray spectra in regions 1 and 2 and list the spectral information in Table~\ref{t:tail}.
The X-ray temperature of region 1 is 1.0 keV, while region 2 is hot at $T \sim$ 4.4 keV, which is discussed in Section~\ref{sec:reg2}.
We can also estimate the mean gas density in the tail from the {\sc apec} normalization.
We assume a uniform density distribution in a cylinder with a height $h \sim 415''$ and a circular base radius $r\sim 46''$ for region 1, and a cylinder with a height $h\sim 163''$ and a circular base radius $r\sim 54''$ for the region 2. The emission measure is $\int n_e n_H dV=n_e^2\frac{n_H}{n_e}\pi r^2h$.
From the Eq.~(\ref{eq:eta}), the mean electron density is 
\begin{equation}
n_e=\sqrt{\frac{4\cdot 10^{14}[D_A(1+z)]^2\eta}{r^2h}\frac{n_e}{n_H}}.  
\end{equation}
The {\sc apec} normalizations are $\eta=(1.8\pm0.6)\times 10^{-4}$ and $\eta=(1.7\pm0.1)\times 10^{-4}$ for regions 1 and 2.
The resultant mean densities are $n_{\rm e,reg1}=(1.2\pm0.2)\times 10^{-3}{\rm\ cm}^{-3}$ and $n_{\rm e,reg2}=(1.6\pm0.1)\times 10^{-3}{\rm\ cm}^{-3}$.
The total hot gas mass of tails are $M_{\rm g,reg1}=(8.3\pm1.4) \times 10^{10} M_{\odot}$ and $M_{\rm g,reg2}=(6.1\pm0.4) \times 10^{10} M_{\odot}$ for regions 1 and 2.

For the detached tail in region 4, we can assume that the gas is roughly distributed within a cone with a height $h\sim 263''$ and a circular base radius $r\sim 110''$. As shown in Fig.~\ref{fig:sbp} (region 4), the SBP of the tail can be fit by a power-law function. 
Thus, we can assume the gas density also follows a power-law function $n_e=ah^p$. We also assume a uniform gas distribution in the circular slice of the cone.
We derive the best-fit parameters of the gas density distribution $n_e=ah^p$ as $a=0.36\pm0.02$ and $p=-1.06\pm0.01$.
Integrating this density distribution over the tail region 4 within the cone, we find a tail mass $M_{\rm g,reg4}=(1.3\pm0.1) \times 10^{11} M_{\odot}$.

\section{Discussion}
\label{sec:discussion}
\subsection{Microphysics of the detached cloud}\label{sec:micro}
The detached X-ray source behind G3 is not a background cluster based on two arguments.
First, its X-ray spectrum shows a strong hump around 0.9 keV, typical of the iron-L hump for $\sim$ 1 keV plasma. If the redshift is left as a free parameter, the 3$\sigma$ upper limit is 0.10 and the best fit agrees with G3 redshift very well.
WINGS provides deep optical imaging data. Within 2$'$ radius of the X-ray peak, there is not a single source that is outside of Z8338 ($z=0.0494$) but luminous enough to be a BCG at $z <$ 0.1.
For the small galaxy close to the X-ray peak (Fig.~\ref{fig:rgb}) to be a BCG, it has to be at $z > 0.8$.
Second, the redshift of a galaxy cluster can also be estimated by aligning the source properties to the known $L_{\rm X} - T$ relation. We force the redshift from 0.05 to 2.0 and fit the X-ray spectrum. The fits get worse with increasing redshift. The best-fit $T$ increases slowly and is still
less than 1.8 keV for $z = 2.0$. If we require the source to fall on the known $L_{\rm X} - T$ relation (e.g. \citealt{Sun2012}), the source redshift needs to be smaller than $\sim$ 0.09.
Thus, we conclude that the X-ray source is not a background cluster, which is consistent with the analysis of \cite{Schellenberger15}. If the X-ray source is in the cluster, its most likely origin is the detached gas halo once associated with G3 group. This is also strongly suggested by the morphology of the X-ray source, which resembles the simulated infalling galaxies with their hot halos being stripped by the ram pressure of the ICM (e.g. \citealt{Roediger2015,Vij17b}).
The stripped halo is composed of a remnant atmosphere and its gas tail and wake (e.g. \citealt{Roediger2015}).
However, in this case, the stripped gas is detached from the host galaxy.

\begin{figure}
\begin{center}
\centering
\includegraphics[angle=0,width=0.49\textwidth]{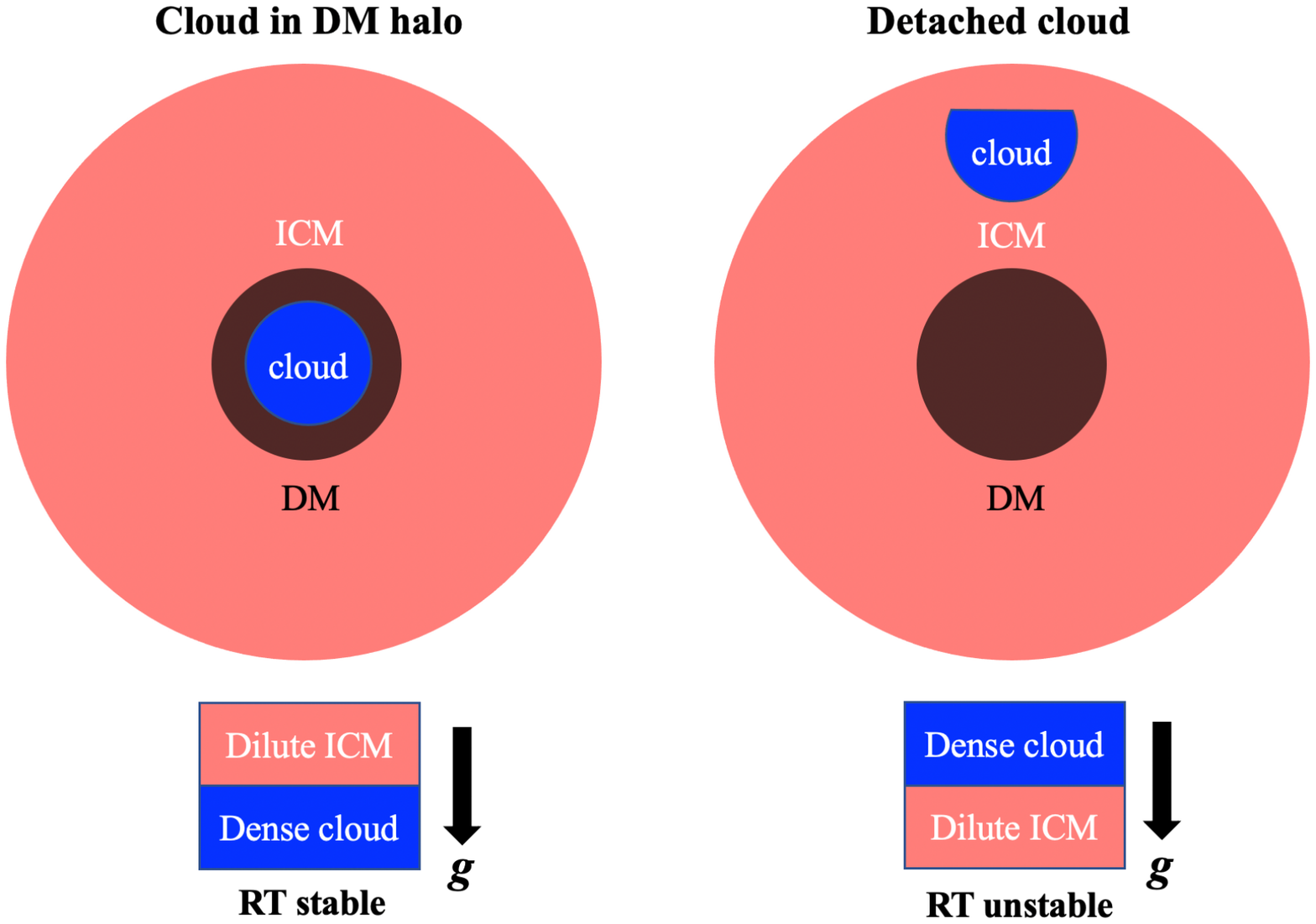}
\caption{
The sketch shows the dependence of the RT instability of the geometry of the detached cloud. When the dense cloud is inside the dark matter (DM) halo, the gravity of the halo can prevent the onset of RT instability. However, if the cloud is stripped into the dilute ICM, the once-associated DM halo further enhances RT instability.
}
\label{fig:rt}
\end{center}
\end{figure}

For a cloud detached during infall, the gravity of the once associated dark matter halo further enhances RT instability as shown in Fig.~\ref{fig:rt}.
RT instability would tear the cloud apart in a few characteristic e-folding times
\begin{equation}
t_{\rm RT} = (\lambda_{\rm RT} / 2 \pi a)^{1/2},
\end{equation}
where $\lambda_{\rm RT}$ is the
wavelength of the RT perturbation and $a=g+a_{\rm d}$ is the effective acceleration
(gravity + drag force combined in this case).
For gravitational acceleration, this detached cloud is influenced by the gravitational pull of the host group and cluster.
The group gravity acceleration $g$ can be estimated assuming an NFW profile ($c_{500}=R_{500}/r_s=4.2$) for the total mass distribution for 1 keV groups (e.g. \citealt{Sun2009}).
We also assume an NFW profile for the cluster ($kT \sim 3$ keV) and estimate the gravitational acceleration at the position of the detached tail as $1.0 \times 10^{-8} {\rm\ cm\ s}^{-2}$. However, the tail alignment is almost perpendicular to the gravity from the cluster. Based on the principle of superposition of forces, the decomposed component of cluster gravity along the tail is insignificant compared with the one from the group gravity.

\begin{figure}
\begin{center}
\centering
\includegraphics[angle=0,width=0.49\textwidth]{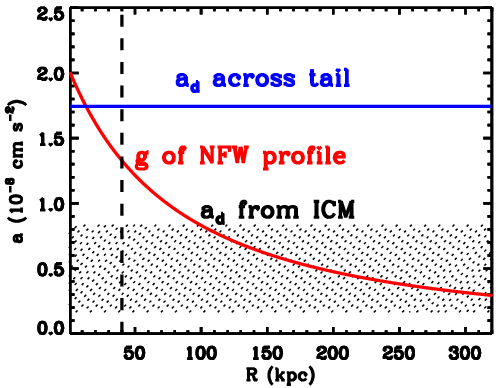}
\caption{
In the comparison between gravity acceleration $g$ and drag acceleration $a_{\rm d}$ due to ram pressure, we need $a_{\rm d}>g$ to remove gas.
The red solid line is the $g$ profile from an NFW total mass profile. The hatched region is the $a_{\rm d}\propto \rho_{\rm ICM}v^2$ from a parameter space for infalling velocity $v$ with surrounding ICM density $\rho_{\rm ICM}$. The blue solid line is the $a_{\rm d}$ across the tail of CC1 with a infalling $v=1300 {\rm\ km\ s}^{-1}$. The dashed line is the currently projected distance of the detached cloud.}
\label{fig:gp}
\end{center}
\end{figure}

The drag force on the detached cloud is $F_{\rm d}=C_{\rm d}\rho_{\rm ICM}v^2A/2$, where $C_{\rm d}\sim 0.4$ is the drag coefficient assuming a hemispherical shape for the detached cloud, the surrounding gas density $\rho_{\rm ICM}=(7.7 \pm 3.9)\times 10^{-28}{\rm\ g\ cm}^{-3}$ with the best-fit density from the $\beta$-model fitting.
The infall velocity $v$ can be constrained from the line of sight velocities of member galaxies. We examined the velocity dispersion of the cluster, with the galaxy velocity values retrieved from NED. Within 29.4$'$ (or 2$R_{500}$) of NGC 6582, there are 98 galaxies with velocities within $3\sigma$ of the
median velocity of the cluster. The median velocity of the cluster galaxies is 14888 km s$^{-1}$, which suggests that BCGs (G1 and G2) may have $\sim -440$ km s$^{-1}$ peculiar velocity. This is not surprising for a cluster in merging.
The line-of-sight velocity dispersion is 723 km s$^{-1}$. The median velocity and velocity dispersion are consistent with results from previous studies (\citealt{Cava2009}), while a rigorous study of the galaxy dynamics and distribution is beyond the scope of this paper.
The group (G3 as its BCG) infall velocity component along the line of sight is then $\sim 470{\rm\ km\ s}^{-1}$ based on the galaxy velocity in Table~\ref{tab_galaxies}. From the surface brightness edge of CC2, the angle constraint of the infall direction with respect to the plane of the sky is $\delta < 35^{\circ}$ (e.g.  \citealt{Mazzotta2001}), otherwise the edge would not be seen in projection if the angle were too large.
The gas tail morphology behind G3 also suggests a substantial velocity component in the plane of the sky. 
If we assume $\delta = 15-35^{\circ}$, the total infall velocity is $v= 1300\pm 500$ km s$^{-1}$.  
The cloud cross-sectional area is $A=\pi r^2$ ($r\sim$ 17 kpc), with a total mass of $M=2.6\times 10^{9} M_\odot$.
We then compare the $g$ with $a_{\rm d}$ in Fig.~\ref{fig:gp}.

We first notice that $g>a_{\rm d}$ at the current position, 40 kpc distant from the group center, marked as a dashed line in Fig.~\ref{fig:gp}. 
Moreover, in the inner region, the drag $a_{\rm d}$ due to ram pressure must overcome the acceleration due to gravity $g$ for ram pressure stripping to remove the gas. 
Therefore, in the past, the ram pressure must be larger than the gravity to remove the gas.
One possibility is that the infalling group has passed through the tail (region 2 in Fig.~\ref{fig:img}) of CC1. The density inside the tail is much higher than nearby ICM. The drag force was boosted significantly when the galaxy was crossing the tail if we assume an infalling $v=1300{\rm\ km\ s}^{-1}$ as shown in Fig.~\ref{fig:gp}. 
In this case, the drag force can strip most of the gas except the very central region where the gravity is larger, thus the central corona can survive from RPS.
The group is infalling supersonic if we assume the same infalling velocity of $v=1300{\rm\ km\ s}^{-1}$ (sound speed of host cluster is $890{\rm\ km\ s}^{-1}$), the shock can also heat the crossed tail of CC1. We have confirmed the high temperature region as in Section~\ref{sec:reg2}.

We adopt average $g$ and $a_{\rm d}$ in Fig.~\ref{fig:gp} to estimate $t_{\rm RT} \sim 16$ Myr ($\lambda_{\rm RT}$ / 17 kpc)$^{1/2}$.
The detached cloud is slowly pushed to the back side of the galaxy so its
average relative velocity to the galaxy should be smaller than the infall velocity
of the galaxy. Even assuming a 1300 km s$^{-1}$ relative velocity, the 40 kpc offset requires 30 Myr.
Thus, RT instability needs to be suppressed here. 
The tangential magnetic field on the
surface of the cloud is an obvious possibility \citep[e.g.][]{Roediger2008RT}.
Magnetic fields perpendicular to the density or temperature gradient are suggested for the suppression of transport processes in the ICM (e.g. \citealt{markevitch2007}; \citealt{Dursi2008}). 
The magnetic field can suppress the growth of perturbations of scale-length $\lambda<\lambda_c$ with
\begin{equation}
\lambda_c=\frac{B^2}{a(\rho_{\rm CLOUD}-\rho_{\rm ICM})}.
\end{equation}
We find a minimum magnetic field $B=6 \mu$G for $\lambda_c=17$ kpc.

When the detached cloud soars through the ICM, there is a velocity difference across the interface between the detached cloud and ICM, and the Kelvin–Helmholtz (KH) instability may develop. 
The magnetic field can also suppress the KH instability if $\frac{B^2(\rho_1+\rho_2)}{2\pi\rho_1\rho_2 (v_1-v_2)^2} \geq 1$ \citep{1961hhs..book.....C}.
In this case, we adopt $v_1-v_2=v=1300{\rm\ km\ s}^{-1}$, $\rho_1=\rho_{\rm CLOUD}$, $\rho_2=\rho_{\rm ICM}$, we get a minimum magnetic field $B=9~\mu$G to suppress the KH instability.
We note that there are relatively large uncertainties in the infalling velocity $v$ and the surrounding ICM density $\rho_{\rm ICM}$, thus the estimation of the magnetic field discussed here only provides a rough order of magnitude estimation.

\subsection{The heated region in the short tail of CC1}\label{sec:reg2}
Double tails trail behind the CC1. 
However, the X-ray temperature of the short tail (Reg2) in Fig.~\ref{fig:img} and Table~\ref{t:src} is significantly higher than the long tail (Reg1), and even higher than the mean cluster temperature.
Their X-ray spectral properties are from the double-background subtraction method as detailed in Section~\ref{sec:data}.
To further check the unusually high temperature of Reg2, we also use the traditional local background subtraction method. 
For the observed data, we have applied the VFAINT filtering that reduces the amplitude of the detector background, and also reduces its gradients and variation across the chip \citep{2006ApJ...645...95H}.
The local background subtraction is mainly subjected to the ICM gradient and exposure vignetting effects.
We produce a mock X-ray image by multiplying the exposure map by the sum of the models for the ICM and sky backgrounds (${\tt [ICM + sky\ model] \times exposure\ map}$) to mimic the ICM gradient and vignetting.
From the mock X-ray image, we select nearby local background regions with similar mean intensity to the short tail (Reg2) in the BI and FI chips, respectively.
We then extract source and local background spectra in the selected regions from different observations.
Jointly fitting the spectra from different chips gives $T_X=3.51\pm0.97$ keV, which is consistent with $T_X=4.44\pm1.13$ keV from the double-background subtraction, after accounting for the errors.

To directly compare the unusually high temperature of Reg2 with Reg1, we assume the X-ray emission in these regions is dominated by the tails, and ignore the ICM or sky backgrounds. We fit the detector background subtracted spectra of Reg1 and Reg2 with a {\sc tbabs*apec} model. 
Although this naive comparison fails to account for the multiple components, it is least subjected to model assumptions. Reg1 and Reg2 are close to each other, we expect that the ICM and sky backgrounds are similar for them, any significant difference from spectra fitting is most likely due to the tail structure.
As shown in Table~\ref{t:tail}, the $T_X$ of Reg2 is still much higher than the one of Reg1. 

The heated Reg2 is probably caused by merger activities like shock heating as discussed in Section~\ref{sec:micro}.
Heated regions between dominant cluster galaxies are observed in other merging clusters (e.g. \citealt{Bogdan2011,su2014}).
Actually, Z8338 contains several substructures that may be merging (e.g. \citealt{Ramella2007}).
Meanwhile, the spectacular double tail feature that stems from CC1 may also be produced by the merger of the subclusters that hosted G1 and G2.
Subcluster merger can further potentially destroy the X-ray CCs (e.g. \citealt{Million2010,Rossetti2011,Ichinohe2015}).

\subsection{Link between the G3 corona and the detached tail}
G3 hosts a radio source with a 1.4 GHz luminosity of
9.7$\times10^{22}$ W/Hz. Radio active galactic nuclei (AGNs) this luminous are almost always associated with an X-ray CC or corona \citep{Sun2009b}.
The gas cooling from the hot phase is able to fuel the radio AGN.
Interestingly, the center of the G3 corona is displaced 2.6$''$ north of the radio AGN or optical center.
The offset suggests that G3 is moving southward, with the displacement caused by the ram pressure of the surrounding ICM.
While the radio AGN remains within the corona (with a radius of 3.7$''$), ram pressure may affect fueling of the AGN.

\begin{figure}
\begin{center}
\centering
\includegraphics[angle=0,width=0.49\textwidth]{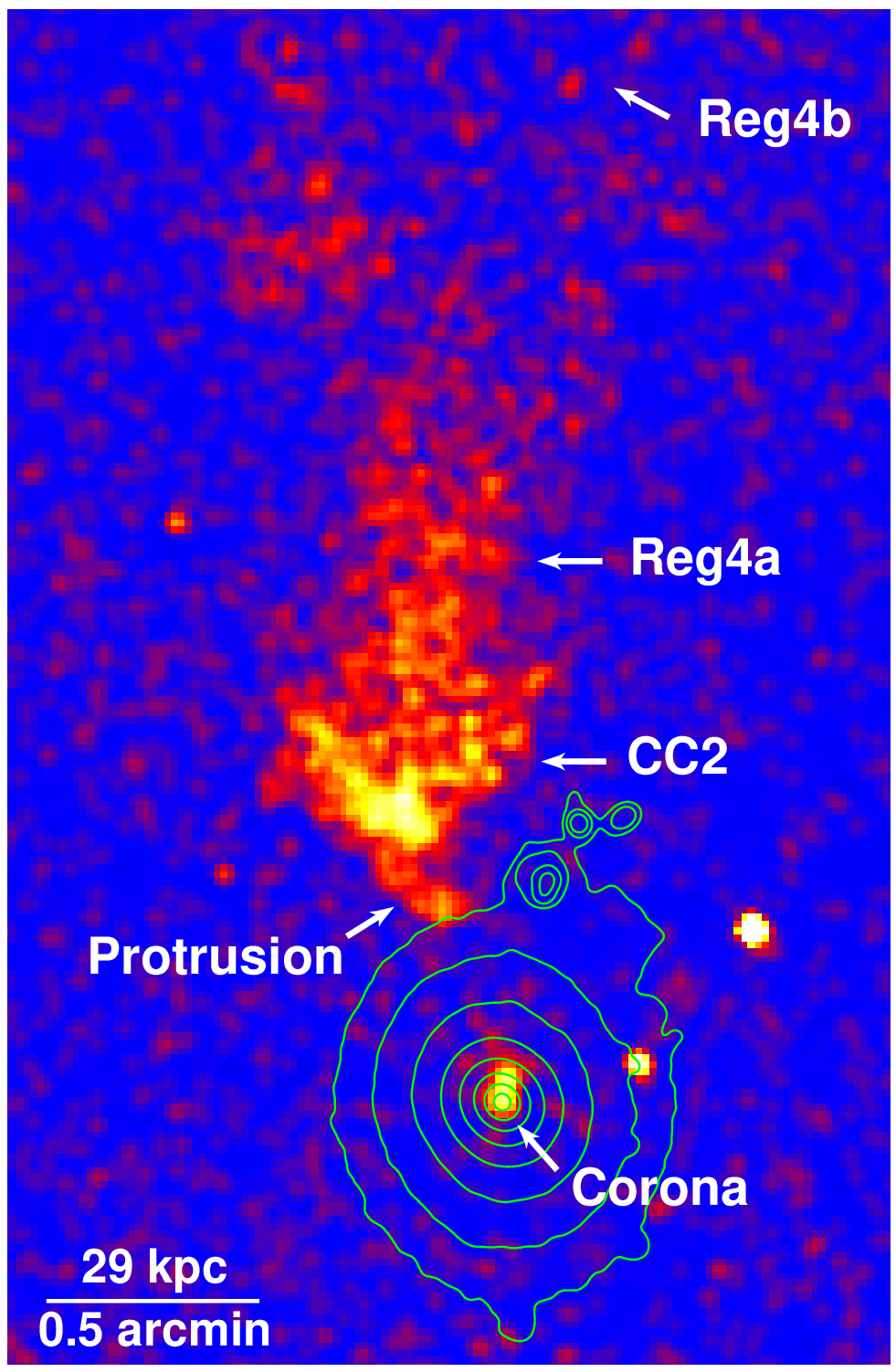}
\caption{
Magnified \cha\ 0.7-2 keV image of the region around the corona of G3 and the detached tail. The green contours are from the WINGS V-band image. The outermost contour approximately matches D25. 
}
\label{fig:zoom}
\end{center}
\end{figure}

\begin{figure}
\begin{center}
\centering
\includegraphics[angle=0,width=0.49\textwidth]{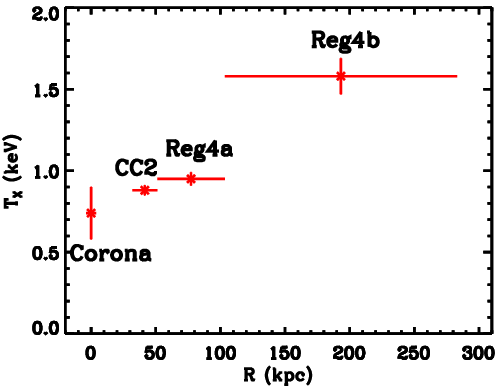}
\caption{
The temperature profile from the G3 corona to the detached tail. The temperatures of corona, detached CC, and remnant tail (Reg4a) are generally consistent with each other, while the temperature of wake (Reg4b) is systematically higher, probably caused by turbulent mixing between the stripped cloud and the hotter ICM.
}
\label{fig:g3tail}
\end{center}
\end{figure}

As shown in region 6 of Figs.~\ref{fig:img} and \ref{fig:sbp}, a protrusion or finger-like part of the detached tail points towards the corona. 
In general, the outer layers of a galaxy's atmosphere are easier to strip than the inner layers, and the outer layers also protect the inner layers until they are stripped.
The protrusion is likely composed of the innermost gas, which was the last to be stripped. 
Meanwhile, the corona extends at least $8''$ from its center towards the protrusion (Fig.~\ref{fig:zoom}). 
The X-ray morphology indicates a link between the detached tail and the corona.
We plot a temperature profile from the corona to the detached tail in Fig.~\ref{fig:g3tail}. The temperature of each region is listed in Tables~\ref{t:src} and \ref{t:tail}. We also mark each region in Fig.~\ref{fig:zoom}. The temperature of the corona is close to those of the remnant atmosphere (CC2) and tail (Reg4a), which suggests the corona is the remainder of the tightly bound gas that has survived RPS. 
Note that the NFW profile near group center in Fig.~\ref{fig:gp} underestimates the true mass from the additional stellar components.
The deep potential in the very center of group can shelter the corona from the ram pressure stripping.

\subsection{Comparison with simulations}

\begin{figure}
\begin{center}
\centering
\includegraphics[angle=0,width=0.49\textwidth]{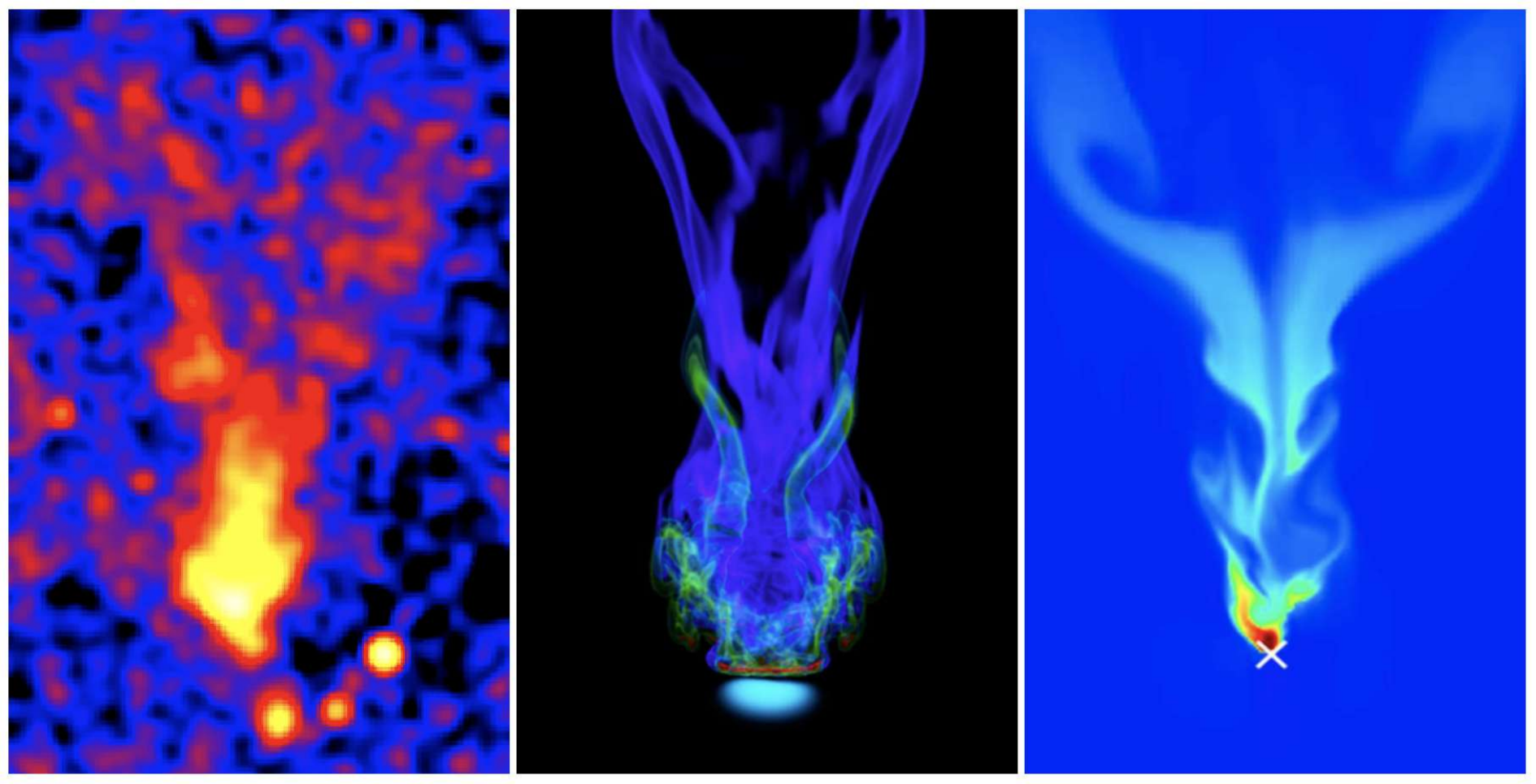}
\vspace{-0.5cm}
\caption{
Compare the observation (left) with MHD simulations from \citealt{Ruszkowski2014} (middle) and \citealt{Vij17b} (right). Note that tails shown from simulations are still attached to the host galaxy.
Although bifurcated tails can be found in purely HD simulations, the presence of magnetic fields makes these tails more stable and less susceptible to shear instabilities.
}
\label{fig:sim}
\end{center}
\end{figure}

The stripped early-type galaxies can be used to probe ICM properties like thermal conductivity, turbulence, viscosity, and magnetic fields (e.g. \citealt{Randall08,su17,Kraft2017}).
The detached tail in Z8338 has a morphology predicted in some simulations mostly for attached tails, a front V-shaped enhancement, a ``trunk'' known as remnant tail behind and narrower than the front, two tails in the wake region split behind the ``trunk''.
Simulations suggest that the delicate structures and morphology of stripped tails are related to ICM microphysics. 
\cite{Roediger2015} studied the effect of viscosity with hydrodynamic (HD) simulations. They found that the viscosity can suppress KH instabilities and mixing, such that viscously stripped galaxies have long X-ray bright cool wakes, with subtle features from KH instabilities suppressed. 
\cite{Shin13} explored the impact of turbulence on ram pressure stripping with HD simulations. They found that galaxies with more turbulent gas produce longer, wide, and more smoothly distributed tails than those characterized by weaker turbulence. Even very weak internal turbulence can significantly accelerate the gas removal from galaxies via RPS. The turbulent motions also help to blend the stripped gas with ambient ICM, as well as alter the composition and metallicity of the stripped tail.
The turbulent mixing between stripped gas and ambient ICM causes an increasing temperature profile as shown in Fig.~\ref{fig:g3tail}.
The higher temperature feature in the wake region was also found by \cite{Schellenberger15}.

\cite{Shin14} further explored the effect of magnetic fields with the magnetohydrodynamic (MHD) simulations.  
The magnetic fields deform the tail morphology significantly. In the tails, the  magnetic field is amplified, with strongly magnetized regions having systematically higher metallicity due to the strong concentration of the stripped gas.
\cite{Ruszkowski2014} also found that magnetic fields have a strong impact on the morphology of the tail. The MHD case shows long filamentary tails, while the purely HD case shows clumpy tails. 
Moreover, the tail bifurcation is due to the general tendency for the magnetic fields to produce filamentary tails with two dominant filaments. 
\cite{Ruszkowski2014} suggested the magnetic fields are stretched along the direction of the ICM wind to form magnetic tails with typical field strengths of a few $\mu$G, and the stripped gas tails are spatially correlated with the magnetic tails.
Although magnetic fields alone are not necessarily responsible for the bifurcated structure as bifurcated tails can be found in the purely HD simulations (e.g. \citealt{Vij15}), the presence of magnetic fields makes these tails more stable and less susceptible to shear instabilities.

We compare the bifurcated tail behind G3 with MHD simulations in Fig.~\ref{fig:sim}. The similarity between the observation and simulations suggests that the tail bifurcation is due to the magnetic field (note that the tails shown from the simulations are still attached to the host galaxy). Double tails are also observed behind other stripped galaxies (e.g. \citealt{Randall08,Sun2010,Zhang2013}).

\section{Conclusions}
\label{sec:conclusion}

In this paper, we present multiple substructures including cool cores, coronae, and tails in a merging galaxy cluster Z8338, mainly based on our \cha\ data.
Our main conclusions are as follows:

(1) A detached ram pressure stripped tail is observed behind the galaxy G3. 
The tail detachment may be caused by the crossing of another tail within region 2 in the middle right panel of Fig.~\ref{fig:img}. The ram pressure is significantly boosted in the higher density tail region and can overcome gravity. 

(2) The other double tail is attached to the CC of G2. Interestingly, the shorter part of this tail ($\sim 200$ kpc, $T_X\sim 4$ keV) is much hotter than the longer part ($\sim 500$ kpc, $T_X\sim 1$ keV), probably due to the shock heating from the merger of the subclusters.

(3) The detached cloud is expected to suffer both RT and KH instabilities, which tend to disintegrate the cloud. We find a few $\mu$G magnetic field can suppress the hydrodynamic instabilities and help the survival of the cloud.

(4) The tail bifurcation may be related to the magnetic fields, which tend to produce filamentary structures  and help to sustain the structures by suppressing the shear instabilities. MHD simulations also show similar bifurcated tails.

(5) A protrusion or finger-like structure at the front tip of the detached tail points to the nucleus of its host galaxy G3, which also hosts a remnant corona extended $\sim$ 8 kpc towards the protrusion. Meanwhile, the temperature of the corona and the innermost detached tail are consistent with each other, suggesting a link between the corona and the detached tail. 
However, the outermost tail shows a systematically higher temperature, suggesting turbulent mixing with the surrounding hotter ICM.

The detached tail or cloud studied here provides us direct insights about CC displacement and disruption due to subcluster merger. The CC, even detached, can still survive for some period of time, which is at least 40 kpc / 1300 ${\rm\ km\ s^{-1}}$ $\sim$ 30 Myr here. The above estimate is certainly a lower limit, as the CC displacement speed is smaller than the infall speed of the subcluster.
The detached tail/cloud also bridges the tails still attached to their hosts and the orphan clouds (e.g. \citealt{Ge2021a}) floating in the intracluster space. 
Although such kinds of detached gas are rarely observed, the partially detached plume near M86 is another example (\citealt{Randall08}), they demonstrate that they can survive for some time after the detachment and affect their surrounding environment (e.g. \citealt{2022A&ARv..30....3B}).
The detached gas is a non-negligible segment in the cluster baryon cycle and may contribute to intracluster clumping, metal, magnetic field, star formation, etc.

\section*{Acknowledgements}

We thank Elke Roediger and the anonymous referee for helpful comments.
This research has made use of the NASA/IPAC Extragalactic Database (NED), which is operated by the Jet Propulsion Laboratory, California Institute of Technology, under contract with the National Aeronautics and Space Administration.
We acknowledge the usage of the HyperLeda database (http://leda.univ-lyon1.fr).
This research has made use of data and/or software provided by the High Energy Astrophysics Science Archive Research Center (HEASARC), which is a service of the Astrophysics Science Division at NASA/GSFC and the High Energy Astrophysics Division of the Smithsonian Astrophysical Observatory.

\section*{Data Availability}
The Chandra raw data used in this paper are available to download at the HEASARC Data Archive website (https://heasarc.gsfc.nasa.gov/docs/archive.html). The reduced data underlying this paper will be shared on reasonable requests to the corresponding authors.




\bibliographystyle{mnras}
\bibliography{z8338}

\newcommand{\noopsort}[1]{}
\begin{thebibliography}{}
\makeatletter
\relax
\def\mn@urlcharsother{\let\do\@makeother \do\$\do\&\do\#\do\^\do\_\do\%\do\~}
\def\mn@doi{\begingroup\mn@urlcharsother \@ifnextchar [ {\mn@doi@}
  {\mn@doi@[]}}
\def\mn@doi@[#1]#2{\def\@tempa{#1}\ifx\@tempa\@empty \href
  {http://dx.doi.org/#2} {doi:#2}\else \href {http://dx.doi.org/#2} {#1}\fi
  \endgroup}
\def\mn@eprint#1#2{\mn@eprint@#1:#2::\@nil}
\def\mn@eprint@arXiv#1{\href {http://arxiv.org/abs/#1} {{\tt arXiv:#1}}}
\def\mn@eprint@dblp#1{\href {http://dblp.uni-trier.de/rec/bibtex/#1.xml}
  {dblp:#1}}
\def\mn@eprint@#1:#2:#3:#4\@nil{\def\@tempa {#1}\def\@tempb {#2}\def\@tempc
  {#3}\ifx \@tempc \@empty \let \@tempc \@tempb \let \@tempb \@tempa \fi \ifx
  \@tempb \@empty \def\@tempb {arXiv}\fi \@ifundefined
  {mn@eprint@\@tempb}{\@tempb:\@tempc}{\expandafter \expandafter \csname
  mn@eprint@\@tempb\endcsname \expandafter{\@tempc}}}

\bibitem[\protect\citeauthoryear{{Asplund}, {Grevesse}, {Sauval}  \&
  {Scott}}{{Asplund} et~al.}{2009}]{Asplund09}
{Asplund} M.,  {Grevesse} N.,  {Sauval} A.~J.,   {Scott} P.,  2009, \mn@doi
  [\araa] {10.1146/annurev.astro.46.060407.145222}, \href
  {http://adsabs.harvard.edu/abs/2009ARA%26A..47..481A} {47, 481}

\bibitem[\protect\citeauthoryear{{Bogd{\'a}n} et~al.,}{{Bogd{\'a}n}
  et~al.}{2011}]{Bogdan2011}
{Bogd{\'a}n} {\'A}.,  et~al., 2011, \mn@doi [\apj]
  {10.1088/0004-637X/743/1/59}, \href
  {https://ui.adsabs.harvard.edu/abs/2011ApJ...743...59B} {743, 59}

\bibitem[\protect\citeauthoryear{{Boselli}, {Fossati}  \& {Sun}}{{Boselli}
  et~al.}{2022}]{2022A&ARv..30....3B}
{Boselli} A.,  {Fossati} M.,   {Sun} M.,  2022, \mn@doi [\aapr]
  {10.1007/s00159-022-00140-3}, \href
  {https://ui.adsabs.harvard.edu/abs/2022A&ARv..30....3B} {30, 3}

\bibitem[\protect\citeauthoryear{{Bower}, {Benson}, {Malbon}, {Helly}, {Frenk},
  {Baugh}, {Cole}  \& {Lacey}}{{Bower} et~al.}{2006}]{Bower06}
{Bower} R.~G.,  {Benson} A.~J.,  {Malbon} R.,  {Helly} J.~C.,  {Frenk} C.~S.,
  {Baugh} C.~M.,  {Cole} S.,   {Lacey} C.~G.,  2006, \mn@doi [\mnras]
  {10.1111/j.1365-2966.2006.10519.x}, \href
  {http://adsabs.harvard.edu/abs/2006MNRAS.370..645B} {370, 645}

\bibitem[\protect\citeauthoryear{{Cava} et~al.,}{{Cava}
  et~al.}{2009}]{Cava2009}
{Cava} A.,  et~al., 2009, \mn@doi [\aap] {10.1051/0004-6361:200810997}, \href
  {https://ui.adsabs.harvard.edu/abs/2009A&A...495..707C} {495, 707}

\bibitem[\protect\citeauthoryear{{Cavaliere} \& {Fusco-Femiano}}{{Cavaliere} \&
  {Fusco-Femiano}}{1976}]{1976A&A....49..137C}
{Cavaliere} A.,  {Fusco-Femiano} R.,  1976, \aap, \href
  {https://ui.adsabs.harvard.edu/abs/1976A&A....49..137C} {500, 95}

\bibitem[\protect\citeauthoryear{{Chandrasekhar}}{{Chandrasekhar}}{1961}]{1961hhs..book.....C}
{Chandrasekhar} S.,  1961, {Hydrodynamic and hydromagnetic stability}

\bibitem[\protect\citeauthoryear{{Dey} et~al.,}{{Dey} et~al.}{2019}]{Dey2019}
{Dey} A.,  et~al., 2019, \mn@doi [\aj] {10.3847/1538-3881/ab089d}, \href
  {https://ui.adsabs.harvard.edu/abs/2019AJ....157..168D} {157, 168}

\bibitem[\protect\citeauthoryear{{Dursi} \& {Pfrommer}}{{Dursi} \&
  {Pfrommer}}{2008}]{Dursi2008}
{Dursi} L.~J.,  {Pfrommer} C.,  2008, \mn@doi [\apj] {10.1086/529371}, \href
  {https://ui.adsabs.harvard.edu/abs/2008ApJ...677..993D} {677, 993}

\bibitem[\protect\citeauthoryear{{Font} et~al.,}{{Font} et~al.}{2008}]{Font08}
{Font} A.~S.,  et~al., 2008, \mn@doi [\mnras]
  {10.1111/j.1365-2966.2008.13698.x}, \href
  {http://adsabs.harvard.edu/abs/2008MNRAS.389.1619F} {389, 1619}

\bibitem[\protect\citeauthoryear{{Ge}, {Wang}, {Burchett}, {Tripp}, {Sun},
  {Li}, {Gu}  \& {Ji}}{{Ge} et~al.}{2018}]{2018MNRAS.481.4111G}
{Ge} C.,  {Wang} Q.~D.,  {Burchett} J.~N.,  {Tripp} T.~M.,  {Sun} M.,  {Li} Z.,
   {Gu} Q.,   {Ji} L.,  2018, \mn@doi [\mnras] {10.1093/mnras/sty2492}, \href
  {https://ui.adsabs.harvard.edu/abs/2018MNRAS.481.4111G} {481, 4111}

\bibitem[\protect\citeauthoryear{{Ge} et~al.,}{{Ge} et~al.}{2021a}]{Ge2021a}
{Ge} C.,  et~al., 2021a, \mn@doi [\mnras] {10.1093/mnras/stab1569}, \href
  {https://ui.adsabs.harvard.edu/abs/2021MNRAS.505.4702G} {505, 4702}

\bibitem[\protect\citeauthoryear{{Ge} et~al.,}{{Ge}
  et~al.}{2021b}]{2021MNRAS.508L..69G}
{Ge} C.,  et~al., 2021b, \mn@doi [\mnras] {10.1093/mnrasl/slab108}, \href
  {https://ui.adsabs.harvard.edu/abs/2021MNRAS.508L..69G} {508, L69}

\bibitem[\protect\citeauthoryear{{Gunn} \& {Gott}}{{Gunn} \&
  {Gott}}{1972}]{Gunn1972}
{Gunn} J.~E.,  {Gott} III J.~R.,  1972, \mn@doi [\apj] {10.1086/151605}, \href
  {http://adsabs.harvard.edu/abs/1972ApJ...176....1G} {176, 1}

\bibitem[\protect\citeauthoryear{{Hickox} \& {Markevitch}}{{Hickox} \&
  {Markevitch}}{2006}]{2006ApJ...645...95H}
{Hickox} R.~C.,  {Markevitch} M.,  2006, \mn@doi [\apj] {10.1086/504070}, \href
  {https://ui.adsabs.harvard.edu/abs/2006ApJ...645...95H} {645, 95}

\bibitem[\protect\citeauthoryear{{Ichinohe}, {Werner}, {Simionescu}, {Allen},
  {Canning}, {Ehlert}, {Mernier}  \& {Takahashi}}{{Ichinohe}
  et~al.}{2015}]{Ichinohe2015}
{Ichinohe} Y.,  {Werner} N.,  {Simionescu} A.,  {Allen} S.~W.,  {Canning}
  R.~E.~A.,  {Ehlert} S.,  {Mernier} F.,   {Takahashi} T.,  2015, \mn@doi
  [\mnras] {10.1093/mnras/stv217}, \href
  {https://ui.adsabs.harvard.edu/abs/2015MNRAS.448.2971I} {448, 2971}

\bibitem[\protect\citeauthoryear{{Jeltema}, {Binder}  \& {Mulchaey}}{{Jeltema}
  et~al.}{2008}]{Jeltema08}
{Jeltema} T.~E.,  {Binder} B.,   {Mulchaey} J.~S.,  2008, \mn@doi [\apj]
  {10.1086/587508}, \href {http://adsabs.harvard.edu/abs/2008ApJ...679.1162J}
  {679, 1162}

\bibitem[\protect\citeauthoryear{{Jones}, {Ryu}  \& {Tregillis}}{{Jones}
  et~al.}{1996}]{Jones96}
{Jones} T.~W.,  {Ryu} D.,   {Tregillis} I.~L.,  1996, \mn@doi [\apj]
  {10.1086/178151}, \href {http://adsabs.harvard.edu/abs/1996ApJ...473..365J}
  {473, 365}

\bibitem[\protect\citeauthoryear{{Karachentsev}}{{Karachentsev}}{1980}]{1980ApJS...44..137K}
{Karachentsev} I.~D.,  1980, \mn@doi [\apjs] {10.1086/190687}, \href
  {http://adsabs.harvard.edu/abs/1980ApJS...44..137K} {44, 137}

\bibitem[\protect\citeauthoryear{{Kraft} et~al.,}{{Kraft}
  et~al.}{2017}]{Kraft2017}
{Kraft} R.~P.,  et~al., 2017, \mn@doi [\apj] {10.3847/1538-4357/aa8a6e}, \href
  {https://ui.adsabs.harvard.edu/abs/2017ApJ...848...27K} {848, 27}

\bibitem[\protect\citeauthoryear{{Machacek}, {Jones}, {Forman}  \&
  {Nulsen}}{{Machacek} et~al.}{2006}]{Machacek06}
{Machacek} M.,  {Jones} C.,  {Forman} W.~R.,   {Nulsen} P.,  2006, \mn@doi
  [\apj] {10.1086/503350}, \href
  {http://adsabs.harvard.edu/abs/2006ApJ...644..155M} {644, 155}

\bibitem[\protect\citeauthoryear{{Markevitch} \& {Vikhlinin}}{{Markevitch} \&
  {Vikhlinin}}{2007}]{markevitch2007}
{Markevitch} M.,  {Vikhlinin} A.,  2007, \mn@doi [\physrep]
  {10.1016/j.physrep.2007.01.001}, \href
  {http://adsabs.harvard.edu/abs/2007PhR...443....1M} {443, 1}

\bibitem[\protect\citeauthoryear{{Mazzotta}, {Markevitch}, {Vikhlinin},
  {Forman}, {David}  \& {van Speybroeck}}{{Mazzotta}
  et~al.}{2001}]{Mazzotta2001}
{Mazzotta} P.,  {Markevitch} M.,  {Vikhlinin} A.,  {Forman} W.~R.,  {David}
  L.~P.,   {van Speybroeck} L.,  2001, \mn@doi [\apj] {10.1086/321484}, \href
  {https://ui.adsabs.harvard.edu/abs/2001ApJ...555..205M} {555, 205}

\bibitem[\protect\citeauthoryear{{McCarthy}, {Frenk}, {Font}, {Lacey}, {Bower},
  {Mitchell}, {Balogh}  \& {Theuns}}{{McCarthy} et~al.}{2008}]{McCarthy08}
{McCarthy} I.~G.,  {Frenk} C.~S.,  {Font} A.~S.,  {Lacey} C.~G.,  {Bower}
  R.~G.,  {Mitchell} N.~L.,  {Balogh} M.~L.,   {Theuns} T.,  2008, \mn@doi
  [\mnras] {10.1111/j.1365-2966.2007.12577.x}, \href
  {http://adsabs.harvard.edu/abs/2008MNRAS.383..593M} {383, 593}

\bibitem[\protect\citeauthoryear{{Million}, {Allen}, {Werner}  \&
  {Taylor}}{{Million} et~al.}{2010}]{Million2010}
{Million} E.~T.,  {Allen} S.~W.,  {Werner} N.,   {Taylor} G.~B.,  2010, \mn@doi
  [\mnras] {10.1111/j.1365-2966.2010.16596.x}, \href
  {https://ui.adsabs.harvard.edu/abs/2010MNRAS.405.1624M} {405, 1624}

\bibitem[\protect\citeauthoryear{{Piffaretti}, {Arnaud}, {Pratt},
  {Pointecouteau}  \& {Melin}}{{Piffaretti} et~al.}{2011}]{Piffaretti2011}
{Piffaretti} R.,  {Arnaud} M.,  {Pratt} G.~W.,  {Pointecouteau} E.,   {Melin}
  J.~B.,  2011, \mn@doi [\aap] {10.1051/0004-6361/201015377}, \href
  {https://ui.adsabs.harvard.edu/abs/2011A&A...534A.109P} {534, A109}

\bibitem[\protect\citeauthoryear{{Ramella} et~al.,}{{Ramella}
  et~al.}{2007}]{Ramella2007}
{Ramella} M.,  et~al., 2007, \mn@doi [\aap] {10.1051/0004-6361:20077245}, \href
  {https://ui.adsabs.harvard.edu/abs/2007A&A...470...39R} {470, 39}

\bibitem[\protect\citeauthoryear{{Randall}, {Nulsen}, {Forman}, {Jones},
  {Machacek}, {Murray}  \& {Maughan}}{{Randall} et~al.}{2008}]{Randall08}
{Randall} S.,  {Nulsen} P.,  {Forman} W.~R.,  {Jones} C.,  {Machacek} M.,
  {Murray} S.~S.,   {Maughan} B.,  2008, \mn@doi [\apj] {10.1086/592324}, \href
  {http://adsabs.harvard.edu/abs/2008ApJ...688..208R} {688, 208}

\bibitem[\protect\citeauthoryear{{Roediger} \& {Hensler}}{{Roediger} \&
  {Hensler}}{2008}]{Roediger2008RT}
{Roediger} E.,  {Hensler} G.,  2008, \mn@doi [\aap]
  {10.1051/0004-6361:200809438}, \href
  {http://adsabs.harvard.edu/abs/2008A%26A...483..121R} {483, 121}

\bibitem[\protect\citeauthoryear{{Roediger} et~al.,}{{Roediger}
  et~al.}{2015}]{Roediger2015}
{Roediger} E.,  et~al., 2015, \mn@doi [\apj] {10.1088/0004-637X/806/1/103},
  \href {https://ui.adsabs.harvard.edu/abs/2015ApJ...806..103R} {806, 103}

\bibitem[\protect\citeauthoryear{{Rossetti}, {Eckert}, {Cavalleri}, {Molendi},
  {Gastaldello}  \& {Ghizzardi}}{{Rossetti} et~al.}{2011}]{Rossetti2011}
{Rossetti} M.,  {Eckert} D.,  {Cavalleri} B.~M.,  {Molendi} S.,  {Gastaldello}
  F.,   {Ghizzardi} S.,  2011, \mn@doi [\aap] {10.1051/0004-6361/201117306},
  \href {https://ui.adsabs.harvard.edu/abs/2011A&A...532A.123R} {532, A123}

\bibitem[\protect\citeauthoryear{{Russell} et~al.,}{{Russell}
  et~al.}{2014}]{Russell14}
{Russell} H.~R.,  et~al., 2014, \mn@doi [\mnras] {10.1093/mnras/stu1469}, \href
  {http://adsabs.harvard.edu/abs/2014MNRAS.444..629R} {444, 629}

\bibitem[\protect\citeauthoryear{{Ruszkowski}, {Br{\"u}ggen}, {Lee}  \&
  {Shin}}{{Ruszkowski} et~al.}{2014}]{Ruszkowski2014}
{Ruszkowski} M.,  {Br{\"u}ggen} M.,  {Lee} D.,   {Shin} M.-S.,  2014, \mn@doi
  [\apj] {10.1088/0004-637X/784/1/75}, \href
  {http://adsabs.harvard.edu/abs/2014ApJ...784...75R} {784, 75}

\bibitem[\protect\citeauthoryear{{Schellenberger} \&
  {Reiprich}}{{Schellenberger} \& {Reiprich}}{2015}]{Schellenberger15}
{Schellenberger} G.,  {Reiprich} T.~H.,  2015, \mn@doi [\aap]
  {10.1051/0004-6361/201527317}, \href
  {http://adsabs.harvard.edu/abs/2015A%26A...583L...2S} {583, L2}

\bibitem[\protect\citeauthoryear{{Shin} \& {Ruszkowski}}{{Shin} \&
  {Ruszkowski}}{2013}]{Shin13}
{Shin} M.-S.,  {Ruszkowski} M.,  2013, \mn@doi [\mnras] {10.1093/mnras/sts071},
  \href {http://adsabs.harvard.edu/abs/2013MNRAS.428..804S} {428, 804}

\bibitem[\protect\citeauthoryear{{Shin} \& {Ruszkowski}}{{Shin} \&
  {Ruszkowski}}{2014}]{Shin14}
{Shin} M.-S.,  {Ruszkowski} M.,  2014, \mn@doi [\mnras]
  {10.1093/mnras/stu1909}, \href
  {http://adsabs.harvard.edu/abs/2014MNRAS.445.1997S} {445, 1997}

\bibitem[\protect\citeauthoryear{{Smith} et~al.,}{{Smith}
  et~al.}{2004}]{Smith04}
{Smith} R.~J.,  et~al., 2004, \mn@doi [\aj] {10.1086/423915}, \href
  {http://adsabs.harvard.edu/abs/2004AJ....128.1558S} {128, 1558}

\bibitem[\protect\citeauthoryear{{Steinhauser}, {Schindler}  \&
  {Springel}}{{Steinhauser} et~al.}{2016}]{Steinhauser2016}
{Steinhauser} D.,  {Schindler} S.,   {Springel} V.,  2016, \mn@doi [\aap]
  {10.1051/0004-6361/201527705}, \href
  {https://ui.adsabs.harvard.edu/abs/2016A&A...591A..51S} {591, A51}

\bibitem[\protect\citeauthoryear{{Su}, {Gu}, {White}  \& {Irwin}}{{Su}
  et~al.}{2014}]{su2014}
{Su} Y.,  {Gu} L.,  {White} Raymond~E. I.,   {Irwin} J.,  2014, \mn@doi [\apj]
  {10.1088/0004-637X/786/2/152}, \href
  {https://ui.adsabs.harvard.edu/abs/2014ApJ...786..152S} {786, 152}

\bibitem[\protect\citeauthoryear{{Su} et~al.,}{{Su} et~al.}{2017}]{su17}
{Su} Y.,  et~al., 2017, \mn@doi [\apj] {10.3847/1538-4357/834/1/74}, \href
  {http://adsabs.harvard.edu/abs/2017ApJ...834...74S} {834, 74}

\bibitem[\protect\citeauthoryear{{Sun}}{{Sun}}{2009}]{Sun2009b}
{Sun} M.,  2009, \mn@doi [\apj] {10.1088/0004-637X/704/2/1586}, \href
  {http://adsabs.harvard.edu/abs/2009ApJ...704.1586S} {704, 1586}

\bibitem[\protect\citeauthoryear{{Sun}}{{Sun}}{2012}]{Sun2012}
{Sun} M.,  2012, \mn@doi [New Journal of Physics]
  {10.1088/1367-2630/14/4/045004}, \href
  {http://adsabs.harvard.edu/abs/2012NJPh...14d5004S} {14, 045004}

\bibitem[\protect\citeauthoryear{{Sun}, {Jones}, {Forman}, {Vikhlinin},
  {Donahue}  \& {Voit}}{{Sun} et~al.}{2007}]{Sun2007}
{Sun} M.,  {Jones} C.,  {Forman} W.,  {Vikhlinin} A.,  {Donahue} M.,   {Voit}
  M.,  2007, \mn@doi [\apj] {10.1086/510895}, \href
  {http://adsabs.harvard.edu/abs/2007ApJ...657..197S} {657, 197}

\bibitem[\protect\citeauthoryear{{Sun}, {Voit}, {Donahue}, {Jones}, {Forman}
  \& {Vikhlinin}}{{Sun} et~al.}{2009}]{Sun2009}
{Sun} M.,  {Voit} G.~M.,  {Donahue} M.,  {Jones} C.,  {Forman} W.,
  {Vikhlinin} A.,  2009, \mn@doi [\apj] {10.1088/0004-637X/693/2/1142}, \href
  {http://adsabs.harvard.edu/abs/2009ApJ...693.1142S} {693, 1142}

\bibitem[\protect\citeauthoryear{{Sun}, {Donahue}, {Roediger}, {Nulsen},
  {Voit}, {Sarazin}, {Forman}  \& {Jones}}{{Sun} et~al.}{2010}]{Sun2010}
{Sun} M.,  {Donahue} M.,  {Roediger} E.,  {Nulsen} P.~E.~J.,  {Voit} G.~M.,
  {Sarazin} C.,  {Forman} W.,   {Jones} C.,  2010, \mn@doi [\apj]
  {10.1088/0004-637X/708/2/946}, \href
  {http://adsabs.harvard.edu/abs/2010ApJ...708..946S} {708, 946}

\bibitem[\protect\citeauthoryear{{Sun} et~al.,}{{Sun} et~al.}{2022}]{Sun2022}
{Sun} M.,  et~al., 2022, \mn@doi [Nature Astronomy]
  {10.1038/s41550-021-01516-8}, \href
  {https://ui.adsabs.harvard.edu/abs/2022NatAs...6..270S} {6, 270}

\bibitem[\protect\citeauthoryear{{Vazza}, {Eckert}, {Simionescu}, {Br{\"u}ggen}
   \& {Ettori}}{{Vazza} et~al.}{2013}]{Vazza2013}
{Vazza} F.,  {Eckert} D.,  {Simionescu} A.,  {Br{\"u}ggen} M.,   {Ettori} S.,
  2013, \mn@doi [\mnras] {10.1093/mnras/sts375}, \href
  {http://adsabs.harvard.edu/abs/2013MNRAS.429..799V} {429, 799}

\bibitem[\protect\citeauthoryear{{Vijayaraghavan} \& {Ricker}}{{Vijayaraghavan}
  \& {Ricker}}{2015}]{Vij15}
{Vijayaraghavan} R.,  {Ricker} P.~M.,  2015, \mn@doi [\mnras]
  {10.1093/mnras/stv476}, \href
  {http://adsabs.harvard.edu/abs/2015MNRAS.449.2312V} {449, 2312}

\bibitem[\protect\citeauthoryear{{Vijayaraghavan} \&
  {Sarazin}}{{Vijayaraghavan} \& {Sarazin}}{2017}]{Vij17b}
{Vijayaraghavan} R.,  {Sarazin} C.,  2017, \mn@doi [\apj]
  {10.3847/1538-4357/aa8bb3}, \href
  {http://adsabs.harvard.edu/abs/2017ApJ...848...63V} {848, 63}

\bibitem[\protect\citeauthoryear{{White} \& {Frenk}}{{White} \&
  {Frenk}}{1991}]{WF91}
{White} S.~D.~M.,  {Frenk} C.~S.,  1991, \mn@doi [\apj] {10.1086/170483}, \href
  {http://adsabs.harvard.edu/abs/1991ApJ...379...52W} {379, 52}

\bibitem[\protect\citeauthoryear{{Willingale}, {Starling}, {Beardmore},
  {Tanvir}  \& {O'Brien}}{{Willingale} et~al.}{2013}]{Willingale13}
{Willingale} R.,  {Starling} R.~L.~C.,  {Beardmore} A.~P.,  {Tanvir} N.~R.,
  {O'Brien} P.~T.,  2013, \mn@doi [\mnras] {10.1093/mnras/stt175}, \href
  {http://adsabs.harvard.edu/abs/2013MNRAS.431..394W} {431, 394}

\bibitem[\protect\citeauthoryear{{Zhang} et~al.,}{{Zhang}
  et~al.}{2013}]{Zhang2013}
{Zhang} B.,  et~al., 2013, \mn@doi [\apj] {10.1088/0004-637X/777/2/122}, \href
  {http://adsabs.harvard.edu/abs/2013ApJ...777..122Z} {777, 122}

\makeatother
\end{thebibliography}

\bsp	
\label{lastpage}
\end{document}